\documentstyle[12pt,epsfig]{article}       
                                                                                              
\newlength{\dinwidth}                                                                                              
\newlength{\dinmargin}                                                                                              
\def\lapproxeq{\lower .7ex\hbox{$\;\stackrel{\textstyle                                                                                              
<}{\sim}\;$}}       
\def\gapproxeq{\lower .7ex\hbox{$\;\stackrel{\textstyle                                                                                              
>}{\sim}\;$}}       
       
\def\be{\begin{equation}}       
\def\ee{\end{equation}}       
\def\bea{\begin{eqnarray}}       
\def\eea{\end{eqnarray}}

\makeatletter                                                                       
\def\fmslash{\@ifnextchar[{\fmsl@sh}{\fmsl@sh[0mu]}}                                                                       
\def\fmsl@sh[#1]#2{%
\mathchoice                                                                       
{\@fmsl@sh\displaystyle{#1}{#2}}%
{\@fmsl@sh\textstyle{#1}{#2}}%
{\@fmsl@sh\scriptstyle{#1}{#2}}%
{\@fmsl@sh\scriptscriptstyle{#1}{#2}}}                                                                       
\def\@fmsl@sh#1#2#3{\m@th\ooalign{$\hfil#1\mkern#2/\hfil$\crcr$#1                                                                       
#3$}}                                                                       
\makeatother        
       
\begin{document}       
       
\titlepage                                                                                              
\begin{flushright}                                                                                              
Budker INP 99-38 \\
DTP/99/38 \\                                                                                              
May 1999 \\                                                                                              
\end{flushright}                                                                                              
       
\vspace*{2cm}       
       
\begin{center}       
{\Large {\bf Infrared safety of impact factors for colourless }} \\       
       
\vspace*{0.3cm}        
{\Large {\bf particle interactions }}       
       
\vspace*{1cm}        
V.S.~Fadin$^{a,b}$ and A.D.~Martin$^b$ \\       
       
\vspace*{0.5cm}        
$^a$ Budker Institute for Nuclear Physics and Novosibirsk       
State University, 630090 Novosibirsk, Russia \\$^b$ Department of Physics,       
University of Durham, Durham, DH1 3LE \\       
\end{center}       
       
\vspace*{2cm}       
\begin{abstract}       
We demonstrate, to next-to-leading order accuracy, the cancellation of the       
infrared singularities in the impact factors which arise in the QCD       
description of high energy processes $A + B \rightarrow A^\prime + B^\prime$       
of colourless particles. We study the example where $A$ is a virtual photon       
in detail, but show that the result is true in general.       
\end{abstract}

\newpage                                                                                             
\noindent {\large \bf 1.~~Introduction}       
       
There continues to be much activity in the study of semihard QCD processes,       
that is processes at small values of $x = Q^2/s$ where $Q^2$ is a typical       
virtuality and $\sqrt{s}$ is the centre-of-mass energy. Many intriguing low $%
x$ phenomena have been observed (see, for example, ref.~\cite{A1}) since the       
announcement of the sharp rise seen in the proton structure function at HERA       
at $x$ decreases. The theoretical understanding of these phenomena is a       
non-trivial task. One important type of process is high energy diffractive $q%
\bar{q}$ (or vector meson) electroproduction, which to lowest order proceeds       
via two gluon exchange. Much data are becoming available for such reactions.       
However the size of the QCD radiative corrections is not known. There are       
indications both experimentally and phenomenologically that these       
corrections are very important \cite{SIZE}. Another relevant process is the total cross  
section for $\gamma^* \gamma^*$ collisions.      
       
The most common basis for the description of small $x$ processes is the BFKL       
equation \cite{LLA}. In this approach the amplitude for the high energy       
process $A+B\rightarrow A^{\prime }+B^{\prime }$ at fixed momentum transfer $%
\sqrt{-t}$ may be symbolically written as the convolution        
\begin{equation}       
\Phi _{A^{\prime }A}\;\otimes \;G\;\otimes \;\Phi _{B^{\prime }B},       
\label{eq:a1}       
\end{equation}       
see Fig.~1 and eq.(\ref{eq:a2}) below. The impact factors    
$\Phi _{A^{\prime}A}$ and $\Phi _{B^{\prime }B}$ describe the transitions    
$A\rightarrow A^{\prime }$ and $B\rightarrow B^{\prime }$ shown by the upper and    
lower blobs of Fig.~1, while $G$ is the Green's function for the two interacting       
Reggized gluons. This representation is valid both in the leading       
logarithmic approximation (LLA) \cite{LLA} , when only the leading terms $%
(\alpha _S\ln s)^n$ are resummed, and in the next-to-leading approximation       
(NLA) \cite{NLA,CC}, when the $\alpha _S(\alpha _S\ln s)^n$ terms are also       
resummed, not only for forward scattering \cite{NLA,CC} with $t=0$, but for       
the non-forward case \cite{NFS} as well.  The explicit form of the kernel   
for forward scattering (with  
colourless exchange in the $t$ channel) was found in ref.~\cite{NLA, CC}.   
A number of subsequent papers (see, for instance, \cite{Ross}) were   
devoted to investigations of its consequences.  
  
Both the kernel of the integral       
equation satisfied by the Green's function and the impact factors are       
unambiguously defined \cite{NLA,NFS,FS2}  in terms of the gluon Regge       
trajectory and the effective vertices for the Reggeon-particle interactions,       
which have been calculated in the NLA in a series of papers   
\cite{NLA2}-\cite{CCH}. \footnote{The impact 
factors were defined in refs.\cite{CC,C,CCC} in a different way; see the comment
  in Section 7.} The cancellation of the infrared   
singularities in the kernel was explicitly demonstrated in ref.~\cite{NLA}.  We       
stress that, for brevity, we are using the terminology ``infrared       
singularities''~to include both soft and collinear singularities.       
Strictly speaking the Green's function for the Reggeized gluons, taken by       
itself, is not physical, since only the amplitudes describing the       
interaction of colourless particles are measurable. It is these physical       
amplitudes that must be free of infrared singularities. However since it has       
been shown that the kernel of the equation for the Green's function is       
infrared safe, it should therefore be possible to construct impact factors which       
are infrared safe. Recall that we are discussing impact factors for       
transitions between colourless particles and, consequently, about a colour       
singlet state in the $t$ channel.       
       
Here we study such impact factors at NLA accuracy. We demonstrate       
explicitly the cancellation of the infrared singularities in the impact       
factor of the virtual photon and, moreover, give a proof that the impact       
factors describing the transitions between colourless particles are infrared       
safe in general. We concentrate on the virtual photon because of the       
importance of the NLA for diffractive $q\bar{q}$ electroproduction, as       
mentioned above. Our study may therefore be considered as a first step in       
analyzing the next-to-leading radiative corrections to this experimentally       
accessible process.       
       
The outline of the paper is as follows. In Section~2 we give a detailed       
definition of the impact factor, concentrating on the example of that for a       
virtual photon. In Section~3 we calculate the singular contribution which       
comes from the {\it virtual} correction to the LLA approximation for the       
effective virtual photon-Reggeon interaction. Then in Section~4 we consider       
the effects arising from an additional {\it real} gluon emitted in the $%
\gamma^*$-Reggeon interaction. In Section~5 we draw these results together       
and demonstrate the cancellation of the infrared singularities in the impact       
factor of the virtual photon in the NLA. A proof of the cancellation in the       
general case is given in Section~6. Finally, in Section~7 we present our       
conclusions. \\       
       
\noindent {\large {\bf 2. The definition of the impact factors}}       
       
First we make explicit the convolution (\ref{eq:a1}) for the amplitude $T$       
describing the process $A + B \rightarrow A^\prime + B^\prime$ shown in       
Fig.~1. To be specific we calculate the $s$-channel imaginary part of the       
amplitude \cite{NFS}        
\begin{eqnarray}  \label{eq:a2}       
{\rm Im} T & = & \int \: \frac{d^{D - 2} q_1}{\mbox{\boldmath $q$}_1^2 %
\mbox{\boldmath $q$}_1^{\prime 2}} \: \int \: \frac{d^{D - 2} q_2}{%
\mbox{\boldmath $q$}_2^2 \mbox{\boldmath $q$}_2^{\prime 2}} \:       
\Phi_{A^\prime A} (\mbox{\boldmath $q$}_1, \mbox{\boldmath $q$}; s_0)        
\nonumber \\       
& & \\       
& & \times \; \frac{s}{(2 \pi)^{D - 2}} \: \int_{\delta - i \infty}^{\delta       
+ i \infty} \: \frac{d \omega}{2 \pi i} \left [ \left ( \frac{s}{s_0} \right       
)^\omega \: G_\omega (\mbox{\boldmath $q$}_1, \mbox{\boldmath $q$}_2; %
\mbox{\boldmath $q$}) \right ] \: \Phi_{B^\prime B} (-\mbox{\boldmath $q$}%
_2, - \mbox{\boldmath $q$}; s_0),  \nonumber       
\end{eqnarray}       
where the momenta are defined in Fig.~1. For convenience we introduce $%
q_i^\prime \equiv q_i - q$, where $q \simeq q_T$ is the momentum transfer in       
the process $A + B \rightarrow A^\prime + B^\prime$        
\begin{equation}  \label{eq:a3}       
q \; = \; p_A - p_{A^\prime} \; = \; p_{B^\prime} - p_B.       
\end{equation}       
$\Phi_{A^\prime A}$ and $\Phi_{B^\prime B}$ are the impact factors which       
describe the $A \rightarrow A^\prime$ and $B \rightarrow B^\prime$ particle       
transitions in the particle-Reggeon scattering processes shown in Fig.~1,       
and $G_\omega$ is the Mellin transform of the Green's function for       
Reggeon-Reggeon scattering. We emphasize that the zig-zag intermediate lines       
in Fig.~1 denote Reggeons, and not gluons --- the Reggeons would only become       
gluons in the absence of interaction. The integrations in (\ref{eq:a2}) are       
performed over $D-2$ dimensional vectors which are transverse to the momenta        
$p_A$ and $p_B$ of the initial particles. The space-time dimension, $D$, is       
taken to be $D = 4 + 2 \varepsilon$ in order to regularize the infrared       
divergences.       
       
Representation (\ref{eq:a2}) is derived assuming that the energy scale $s_0$       
is independent of $\mbox{\boldmath $q$}_1$ and $\mbox{\boldmath $q$}_2$, or       
at least has a factorizable dependence on $\mbox{\boldmath $q$}_1$ and $%
\mbox{\boldmath $q$}_2$. Then any $s_0$ dependence induced by using $s_0$ as       
the energy scale in $G_\omega$ can, to NLA accuracy, be transferred to the       
impact factors. This was shown in ref.~\cite{A9} for forward scattering; for       
the non-forward case the proof is essentially unchanged.       
       
We consider a colour singlet state in the $t$-channel (or, to be more precise, vacuum  
or Pomeron quantum number exchange), since only in this case will the infrared       
singularities cancel. Thus we can suppress the colour group representation       
indices of the impact factors and the Green's function. We also note that       
representation (\ref{eq:a2}) applies not only to elastic processes, but also       
to reactions where $A^{\prime }$ and $B^{\prime }$ may each correspond to a       
group of particles with invariant mass independent of $s$. Finally the       
normalisation adopted in (\ref{eq:a2}) is that of \cite{NFS}, which seems to       
be convenient for non-forward processes. It differs at $t=0$ from that used       
previously such that        
\begin{equation}       
\left. G_\omega (\mbox{\boldmath $q$}_1,\mbox{\boldmath $q$}_2;%
\mbox{\boldmath $q$})\right| _{\mbox{\boldmath $q$}=0}\;=\;%
\mbox{\boldmath                   
$q$}_1^2\mbox{\boldmath $q$}_2^2G_\omega (\mbox{\boldmath $q$}_1,%
\mbox{\boldmath $q$}_2)  \label{eq:a4}       
\end{equation}       
where $G_\omega (\mbox{\boldmath $q$}_1,\mbox{\boldmath $q$}_2)$ is the       
Mellin transform that was used for the forward case \cite{NLA}.       
       
We may express both the kernel of the equation for the Green's function and       
the impact factors in terms of the gluon Regge trajectory and the effective       
vertices for the Reggeon-particle interaction \cite{NFS}. We begin with the       
high energy amplitude $T^{(8)}$ for the process $A+B\rightarrow a+b$ of Fig.~2,   
which is an intermediate process of our reaction $%
A+B\rightarrow A^{\prime }+B^{\prime }$. We have colour octet Reggeized       
gluon exchange in the $t_R$ channel $(t_R=q_R^2\simeq \mbox{\boldmath $q$}%
_R^2)$ with negative signature so that        
\begin{equation}       
T^{(8)}\;=\;\frac{2s}{t_R}\:\Gamma _{aA}^i(q_R)\left[ \frac 12\left( \frac{-s%
}{-t_R}\right) ^{\omega (t_R)}\:+\frac 12\left( \frac s{-t_R}\right) ^{\omega       
(t_R)}\right] \:\Gamma _{bB}^i(-q_R),  \label{eq:a5}       
\end{equation}       
where $\Gamma _{aA}^i$ and $\Gamma _{bB}^i$ are the Reggeon vertices for       
transitions $A\rightarrow a$ and $B\rightarrow b$ and $1\:+\omega (t)$ is the       
gluon Regge trajectory.  Note that this representation is true for       
particle-parton as well as for parton-parton processes. By parton we mean a       
coloured quark or gluon, whereas a particle is colourless. In the       
parton-parton case the vertices are known in the one-loop approximation       
\cite{VER}. One of our aims is to identify the particle-parton vertices in       
the same approximation. The trajectory function,        
\begin{equation}       
j(t)\;=\;1\:+\:\omega (t),  \label{eq:a6}       
\end{equation}       
is independent of the nature of the scattering  
 objects, and is now known in       
the two-loop approximation \cite{TRA}. The summation over the repeated       
index $i$ is over the 8 coloured gluon states.       
       
In the LLA the impact factors are given by       
\begin{equation}       
\Phi _{A^{\prime }A}^{(0)}(\mbox{\boldmath $q$}_R,\mbox{\boldmath $q$}%
)\;=\;\frac 1{\sqrt{N_C^2-1}}\:\sum_{\{a\}}\:\int \:\frac{dM_a^2}{2\pi }\:\Gamma       
_{aA}^{(0)i}(q_R)\:\Gamma _{aA^{\prime }}^{(0)i}(q_R^\prime)^*\:d\rho _a         
\label{eq:a7}       
\end{equation}       
where the number of colours $N_C=3$ for QCD and $\Gamma ^{(0)}$ means that       
the vertices are evaluated in the LLA or Born approximation. The sum $\{a\}$       
is over all states $a$ which contribute to the $A\rightarrow A^{\prime }$       
transition, and over all discrete quantum numbers of these states. The phase       
space element $d\rho _a$ of a state $a$ consisting of particles with momenta        
$\ell _n$ is        
\begin{equation}       
d\rho _a\;=\;(2\pi )^D\:\delta \left( p_A-q_R-\sum_{n\epsilon a}\ell       
_n\right) \:\prod_{n\epsilon a}\:\frac{d^{D-1}\ell _n}{(2\pi )^{D-1}2E_n},       
\label{eq:a8}       
\end{equation}       
while the remaining integration in (\ref{eq:a7}) is over the squared       
invariant mass of the state $a$        
\[       
M_a^2\;=\;(p_A-q_R)^2\;=\;(p_A^{\prime }-q_R^{\prime })^2.        
\]       
In the LLA the impact factors do not depend on $s_0$. In fact the energy       
scale is arbitrary in this approximation.       
       
To be specific, we choose the relevant example of the impact factor of a       
virtual photon $\gamma^*$. In order to define the vertices $%
\Gamma_{a\gamma^*}^i$ we may study the high energy interaction of the $%
\gamma^*$ with any particle. Clearly we should take the simplest choice for       
particle $B$ --- a massless quark with momentum $p_B = p_2$ and $p_2^2 = 0$       
--- and introduce the light-cone momentum $p_1$ so that the virtual photon       
momentum is        
\begin{equation}  \label{eq:a9}       
p_A \; = \; p_1 \: - \: \frac{Q^2}{s} p_2 \quad\quad {\rm and} \quad\quad s       
\; = \; 2 p_1 . p_2,       
\end{equation}       
where $- Q^2$ is the virtuality of the photon.       
       
The lowest-order $\gamma ^{*}q$ process which has a non-vanishing cross       
section at large $s$ is $\gamma ^{*}q\rightarrow (q\bar q)q$ where the $%
q\bar q$ pair is produced in the fragmentation region of the photon. The       
Feynman diagrams are shown in Fig.~3. We may evaluate these diagrams using       
the trick of writing the metric tensor in the gluon propagator as        
\begin{equation}       
g^{\mu \nu }\;=\;\frac{2p_2^\mu p_1^\nu }s\:+\:\frac{2p_1^\mu p_2^\nu }%
s\:+\:g_{\perp }^{\mu \nu },  \label{eq:a10}       
\end{equation}       
and retaining just the first term. Only this term gives a leading (growing       
as $s$) contribution to the amplitude. In this way we arrive at the $\gamma       
^{*}q\rightarrow (q\bar q)q$ Born amplitude in the form of (\ref{eq:a5})       
with the lower vertex (describing the $q\rightarrow q^{\prime}$       
transition via the Reggeon field) given by       
\begin{equation}       
\Gamma _{q^{\prime },q}^{(0)i}\;=\;g\bar u(p_2^{\prime })\:\frac{       
\fmslash {p}_1}{s}\:t^i\:u(p_2)\;=\;g\langle       
q^{\prime }|t^i|q\rangle \:\delta _{\lambda _2\lambda _2^{\prime }},       
\label{eq:a11}       
\end{equation}       
and the upper vertex (describing the $\gamma ^{*}\rightarrow q\bar q$       
transition) given by       
\begin{equation}       
\Gamma _{q\bar q,\gamma ^{*}}^{(0)i}(q_R)\;=\;       
\langle q|t^i|\bar q\rangle \Gamma       
_{q\bar q,\gamma ^{*}}^{}(q_R);\:\:\:\Gamma _{q\bar q,\gamma       
^{*}}(q_R)=\Gamma _a \:+\:\Gamma _b       
\label{eq:a12}       
\end{equation}       
with        
\begin{equation}       
\Gamma _a^{} \:=\:\frac{ee_qg}s\:\bar u(\ell _{+})\:       
\frac{\fmslash {\varepsilon }_{\gamma ^{*}}(\fmslash {\ell }_{-}+       
\fmslash{q}_R)\fmslash {p}_2}{\Delta _{-}}\:v(\ell _{-})        
\label{eq:a13}       
\end{equation}       
\begin{equation}       
\Gamma _b^{} \:=\:-\:\frac{ee_qg}s\:\bar u(\ell _{+})\:\frac{\fmslash {p}_2       
(\fmslash{\ell }_{+}+%
\fmslash{q}_R)\fmslash{\varepsilon }_{\gamma ^{*}}}{\Delta _{+}}\:v(\ell _{-}).       
\label{eq:a14}       
\end{equation}       
The components $\Gamma _a$ and $\Gamma _b$ correspond to the upper vertices       
in diagrams 3a and 3b respectively. The notation used in        
(\ref{eq:a11})--(\ref{eq:a14}) is as follows: $\lambda $ denotes the quark       
helicity, $g$ is the QCD coupling and $\langle a|t^i|b\rangle $ are the       
matrix elements of the colour generator in the fundamental representation.       
Also $ee_q$ is the electric charge of the quark (in the $q\bar q$ pair), $%
\varepsilon _{\gamma ^{*}}$ is the polarization vector of the photon, $%
\ell _{\pm }$ are the $q$ and $\bar q$ momenta shown in Fig.~3 and, finally,        
\begin{equation}       
\Delta _{\pm }\;=\;\frac{\mbox{\boldmath $\ell$}_{\mp }^2}{x_{\mp }}%
\:+\:x_{\pm }Q^2.  \label{eq:a15}       
\end{equation}       
We have taken the quark to be massless and have used the decomposition       
\begin{equation}       
\ell _{\pm }\;=\;x_{\pm }p_1+\frac{\mbox{\boldmath $\ell$}_{\pm }^2}{sx_{\pm       
}}p_2\:+\:\ell _{\pm \perp },\:\:\:\mbox{\boldmath $\ell$}_{\pm }^2       
\equiv -\ell_{\pm \perp }^2.  \label{eq:a16}       
\end{equation}       
In fact the expressions (\ref{eq:a11})--(\ref{eq:a14}) for the vertices       
could have been written immediately by noting that the vertex factors%
\footnote{These are the simplest vertices in the effective action for Reggeon       
interactions \cite{A10} which contains all LLA Reggeon-parton vertices.} $%
igt^i\fmslash{p}_{1,2}/s$ describe the       
interaction of a Reggeon with quarks having a large component of momentum       
along $p_{2,1}$ respectively. In other words in the LLA the Reggeon acts as       
gluon with polarization vector $-p_{2,1}/s$ when it interacts with a quark       
having a large component of momentum along $p_{1,2}$ respectively. Taking       
this into account we can represent the vertices $\Gamma _{q\bar q;       
\gamma^{*}}^{(0)i}$ by the diagrams of Fig.~4.       
       
The vertex $\Gamma _{q\bar q;\gamma ^{*\:^{\prime }}}^{(0)i}(q_R^{\prime})$        
can be calculated in just the same way.  It is given by (\ref{eq:a12})--(\ref{eq:a14})       
with replacements $\gamma ^{*}\rightarrow \gamma ^{*\:^{\prime }},$ $%
q_R\rightarrow q_R^{^{\prime }},$ $\Delta _{\pm }\rightarrow \Delta _{\pm       
}^{\prime },$ where       
\begin{equation}       
\Delta _{\pm }^{^{\prime }}\;=\;\frac{(\mbox{\boldmath $\ell$}_{\mp }+x_{\mp       
}\mbox{\boldmath $q$})^2}{x_{\mp }}\:+\:x_{\pm }Q^{^{\prime }\:2},       
\label{eq:a15'}       
\end{equation}       
and $-Q^{\prime 2}$ is the virtuality of the final photon.       
       
Let us denote  
\begin{equation}  
\frac 1{\sqrt{N_C^2-1}}\sum_{\{q\bar q\}}\:\Gamma _{q\bar q;\gamma  
^{*}}^{(0)i}(q_R)\:\Gamma _{q\bar q;\gamma ^{* \prime}}^{(0)i}  
(q_R^{\prime})^*=f_{q\bar q}(x_{\pm }, \mbox{\boldmath $\ell$}_{\pm },   
\mbox{\boldmath $q$}_R, \mbox{\boldmath $q$}_R^{\prime}).  
\label{eq:a16a}  
\end{equation}  
Besides the variables explicitly shown in (\ref{eq:a16a}),   
the function $f_{q\bar q}$ also depends on the fixed parameters $%
Q^2,\:Q^{^{\prime }\:2}$ and on the photon polarisation  
vectors. The photon impact factor in the NLA is given by the integral of  
the function $f_{q\bar q}$ over  
\[  
\frac{dM_{q\bar q}^2}{2\pi }\:d\rho _{q\bar q}\;=2\delta  
(1-x_{+}-x_{-})(2\pi )^{D-1}\delta (\mbox{\boldmath $\ell$}_{+}+%
\mbox{\boldmath $\ell$}_{-}+\mbox{\boldmath $q$}_R)  
\]  
\begin{equation}  
\times \frac{dx_{+}}{2x_{+}}\:\frac{dx_{-}}{2x_{-}}\frac{d^{D-2}\ell   
_{+\perp}}{%
(2\pi )^{D-1}}\frac{d^{D-2}\ell _{-\perp}}{(2\pi )^{D-1}}.  \label{eq:a17}  
\end{equation}  
In fact this impact factor differs by only trivial factors from the  
analogous one in QED, which was obtained for some particular cases many years  
ago \cite{GLF}.  The explicit form of the function $f_{q\bar q}$ is not   
important for our purposes. The crucial properties of this function are:   
\begin{equation}  
f_{q\bar q}(x_{\pm }, \mbox{\boldmath $\ell$}_{\pm },   
\mbox{\boldmath $0$}, \mbox{\boldmath $q$}_R^{\prime})=  
f_{q\bar q}(x_{\pm }, \mbox{\boldmath $\ell$}_{\pm },   
\mbox{\boldmath $q$}_R, \mbox{\boldmath $0$})=0.   
\label{eq:a17a}  
\end{equation}  
These properties can be verified directly by using expressions   
(\ref{eq:a12})-(\ref{eq:a14}) for $\Gamma _{q\bar q;\gamma  
^{*}}^{(0)i}(q_R)$  and corresponding expressions for   
$\Gamma _{q\bar q;\gamma ^{* \prime}}^{(0)i}  
(q_R^{\prime})$. But, in fact, these properties have a quite   
general nature and are consequences of gauge invariance.   
These properties guarantee that the impact factors of colourless   
objects satisfy the equations  
\begin{equation}  
\Phi_{A^{\prime}A}(\mbox{\boldmath $0$},\,\mbox{\boldmath $q$})  
=\Phi_{A^{\prime}A}(\mbox{\boldmath $q$},\,\mbox{\boldmath $q$})=0,   
\label{eq:a17b}  
\end{equation}  
that are necessary for the corresponding cross sections to be finite.  The next   
important fact is that the integral of the function $f_{q\bar q}$ over the measure   
(\ref{eq:a17}) is not singular, so that we can take $x_{\pm }\sim 1,\:$   
$\mbox{\boldmath $\ell$}_{\pm }^2$ $\sim \mu^2, $ with some nonzero $\mu $   
depending on $Q^2,\:Q^{^{\prime }\:2}, \mbox{\boldmath   
$q$}_R^2,\:(\mbox{\boldmath $q$}_R-\mbox{\boldmath $q$})_{}^2. $ For our   
purposes we can consider all these parameters to be of the same order. In the   
following we use $\mu ^2$ to denote this order.  
      
In the NLA the expression (\ref{eq:a7}) for the impact factor has to be       
changed in two ways. First we have to take into account the radiative       
corrections to the vertices $\Gamma $ considered above. We note that the       
corrections depend on an energy scale. In the definition (\ref{eq:a5}) the       
energy scale was put equal to $-t_R$. Since we calculate impact factors for       
scale $s_0$ we must allow for the change of the vertices arising from the       
change of scale. It is clear from (\ref{eq:a5}) that the ratio of vertex       
factors evaluated at scales $s_1$ and $s_2$ is equal to $(s_2/s_1)^{\omega       
(t_R)/2}$ for a Reggeon of momentum $q_R$. Secondly, in the sum over $\{a\}$       
in (\ref{eq:a7}), we have to include more complicated states which appear in       
the next order of perturbative theory, such as states with an additional       
gluon. For the photon impact factor $q\bar qg$ is the only such additional       
state. But here we have a problem. The integral over $M_a^2$, which is well       
convergent in the LLA case, becomes divergent when an extra gluon is       
included in the final state. The divergence arises because the gluon may be       
emitted not only in the fragmentation region of particle $A$, but also in       
the central region and in the fragmentation region of particle $B$. All       
three contributions have to be carefully separated and the last two assigned       
to the Green's function $G$ and the impact factor $\Phi _{B^{\prime }B}$       
respectively. The separation was described in ref.~\cite{FS2}. The result       
for the non-forward case is \cite{NFS}       
\begin{eqnarray}       
\label{eq:a19}   
\Phi _{A^{\prime }A}(\mbox{\boldmath $q$}_R,\mbox{\boldmath $q$};s_0)       
&=&\frac 1{\sqrt{N_C^2-1}}\:\sum_{\{a\}}\:\int \:\frac{dM_a^2}{2\pi }\:\Gamma       
_{aA}^i(q_R)\:\Gamma _{aA^{\prime }}^{i}(q_R^{\prime})^*\:d\rho _a\theta        
(s_\Lambda -M_a^2)        
\nonumber \\       
&&  \nonumber \\       
&&-\;\frac 12\:\int \:\frac{d^{D-2}q_{R^{\prime }}}{\mbox{\boldmath        
$q$}_{R^{\prime }}^2\mbox{\boldmath $q$}_{R^{\prime }}^{\prime 2}}\:\Phi       
_{A^{\prime }A}^{(0)}(\mbox{\boldmath $q$}_{R^{\prime }},%
\mbox{\boldmath        
$q$})\:\left[ K_r^{(0)}(\mbox{\boldmath $q$}_{R^{\prime }},%
\mbox{\boldmath        
$q$}_R;\mbox{\boldmath $q$})\:\ln \left( \frac{s_\Lambda ^2}{(%
\mbox{\boldmath        
$q$}_{R^{\prime }}-\mbox{\boldmath $q$}_R)^2s_0}\right) \right. \\       
&&  \nonumber \\       
&&+\;\mbox{\boldmath $q$}_{R^{\prime }}^{\prime 2}\mbox{\boldmath        
$q$}_{R^{\prime }}^2\:\delta (\mbox{\boldmath $q$}_R-\mbox{\boldmath        
$q$}_{R^{\prime }})\:\left. \left\{ \omega (t_R)\:\ln \left( \frac{%
\mbox{\boldmath        
$q$}_R^2}{s_0}\right) \:+\:\omega (t_R^{\prime })\:\ln \left( \frac{%
\mbox{\boldmath $q$}_R^{\prime 2}}{s_0}\right) \right\} \right] ,  \nonumber           
\end{eqnarray}       
where the intermediate parameter $s_\Lambda $ should go to infinity. The       
second term on the right hand side of (\ref{eq:a19}) is the subtraction       
of the contribution of the emission of the additional gluon outside the fragmentation       
region of particle $A$.  The dependence on $s_\Lambda$ vanishes because of the   
cancellation between the first and second terms.  $K_r^{(0)}$ is that part of LLA   
kernel of the equation for the non-forward Green's function which is related to real  
gluon production \cite{LLA}  
\begin{equation}       
\label{eq:a20}   
K_r^{(0)}(\mbox{\boldmath $q$}_{R^{\prime }},\mbox{\boldmath $q$}_R;%
\mbox{\boldmath $q$})\;=\;\frac{g^2N_C}{(2\pi )^{D-1}}\left( \frac{%
\mbox{\boldmath $q$}_R^2\mbox{\boldmath $q$}_{R^{\prime }}^{\prime 2}+%
\mbox{\boldmath $q$}_R^{\prime 2}\mbox{\boldmath               
$q$}_{R^{\prime }}^2}{(\mbox{\boldmath $q$}_R-\mbox{\boldmath               
$q$}_{R^{\prime }})^2}-\mbox{\boldmath $q$}^2\right) .    
\end{equation}       
The kernel describes the transition from the $t$ channel state of two       
Reggeons with momenta $q_R$ and $-q_R^{\prime }\equiv q-q_R$ into the state       
with $q_{R^{\prime }}$ and $-q_{R^{\prime }}^{\prime }\equiv q-q_{R^{\prime       
}}$. The terms which involve the Regge trajectory in (\ref{eq:a19}) arise       
from the change of energy scale in the vertices $\Gamma $. The trajectory       
function $\omega (t)$ has to be taken in the one loop approximation \cite       
{LLA}        
\be  
\omega (t)\;=\;\frac{g^2t}{(2\pi )^{D-1}}\frac{N_C}2\int \frac{d^{D-2}k}{%
\mbox{\boldmath $k$}^2(\mbox{\boldmath $q$}-\mbox{\boldmath       
$k$})^2}\;=\;-\bar g^2(\mbox{\boldmath $q$}^2)^\epsilon \frac{\Gamma       
^2(\epsilon )}{\Gamma (2\epsilon )}, \label{eq:b20}    
\ee  
where $\Gamma (x)$ is the Euler function and        
\begin{equation}       
\bar g^2\;\equiv \;g^2N_C\Gamma (1-\epsilon )/(4\pi )^{D/2}.       
\label{eq:a21}       
\end{equation}       
       
As stated above, to evaluate the photon impact factor in NLA we have to       
evaluate the virtual corrections to the $\gamma ^{*}\rightarrow q\bar q$       
vertex of (\ref{eq:a12}) and to include the effects of the emission of an       
additional gluon (that is to find the $\gamma ^{*}\rightarrow q\bar qg$       
vertices and to include their contribution in (\ref{eq:a19})). We begin       
with the virtual corrections. \\       
       
\noindent {\large {\bf 3.~~One-loop corrections to the $\gamma^* R       
\rightarrow q\bar{q}$ vertex}}       
       
One way to determine the corrections to the $\gamma^* R \rightarrow q\bar{q}$         
vertex is to calculate the one-loop corrections to the diagrams of Fig.~3 with a    
colour-octet state          
in the $t_R$ channel.  We keep not only the $\ln s$ terms, but the constant terms as          
well.  We will find the coefficient of the $\ln s$ term, which is due to gluon          
Reggeization, is       
\be         
\label{eq:a22}         
\frac{2s}{t_R} \: \Gamma_{q\bar{q}, \gamma^*}^{(0) i} \:        
\Gamma_{q^{\prime},q}^{(0) i} \:          
\omega (t_R),         
\ee        
as it must be.  If we compare the calculated corrected amplitude with representation          
(\ref{eq:a5}) and use the known\footnote{It was determined using the same procedure          
for the quark-quark scattering process.} expression for the lower vertex        
$\Gamma_{q^{\prime},q}^i$ to one-loop accuracy \cite{VER}, then we may extract          
$\Gamma_{q\bar{q}, \gamma^*}^i$ in the one-loop approximation.         
         
In practice we can determine the $\gamma^* R \rightarrow q\bar{q}$ vertex to          
one-loop accuracy in a more economical way by focusing attention directly on the          
vertex and not on the amplitude.  To be specific we work in the Feynman gauge.           
There are several ways in which we can add the extra virtual gluon line to the          
diagrams of Fig.~3.  It is clear that if one end of the extra gluon line is attached to the          
$q$ or $\bar{q}$ of the upper $q\bar{q}$ pair and the other end is either again          
attached to one of the upper pair or to the exchanged gluon, then the correction is to          
the upper vertex.  Similar considerations apply to the lower vertex.  It is also clear that          
half the gluon self-energy correction has to be included in the upper vertex and half in          
the lower vertex.  The only uncertainty comes from the diagrams of Fig.~5 with two          
gluon lines exchanged between the upper and the lower vertices.  Perhaps a better          
terminology, which we shall adopt, is to say two gluons exchanged between the upper         
and          
lower blocks.  We analyse these two-gluon exchange diagrams first.         
           
\noindent {\bf 3.1~~Two-gluon exchange diagrams}         
         
We start with diagram 5(a).  The basic integral for this diagram is         
\be         
\label{eq:a23}         
I_{5a} \; = \; \int \: \frac{d^D k}{(2 \pi)^D i} \: \frac{1}{(k^2 + i0) ((q_R - k)^2 + i 0)          
((p_2^\prime - k)^2 + i 0) ((\ell_- + k)^2 + i 0)}.         
\ee         
It is convenient to use the Sudakov decomposition for the gluon momentum,  
\be         
\label{eq:a24}         
k \; = \; \beta p_1 + \alpha p_2 + k_\perp, \quad\quad d^{D} k \; = \; \frac{s}{2} \: d          
\alpha d \beta d k_\perp^{D-2},         
\ee         
and to consider three regions of the variables $\alpha, \beta$ :         
\bea         
\label{eq:a25}         
{\rm the~central~region} & & | \alpha | \; < \; \alpha_0, \quad | \beta | \; < \;          
\beta_0 \nonumber \\         
{\rm the}~A~{\rm region} & & | \beta | \; \geq \; \beta_0 \\         
{\rm the}~B~{\rm region} & & | \alpha | \; \geq \; \alpha_0, \nonumber         
\eea         
with $\alpha_0$ and $\beta_0$ chosen so that       
\be         
\label{eq:a26}         
\alpha_0 \; \ll \; 1, \quad \beta_0 \; \ll \; 1 \quad\quad {\rm and} \quad\quad s \alpha_0          
\beta_0 \; \gg \; \mu^2.         
\ee         
Recall that we use $\mu$ to denote the order of magnitude of typical transverse        
momenta.  In fact we will take the limit $s \alpha_0 \beta_0 \rightarrow \infty$, while        
$\alpha_0          
\rightarrow 0$ and $\beta_0 \rightarrow 0$.  We see that regions $A$ and $B$         
overlap.  The          
whole $\alpha, \beta$ plane is covered by the sum of the three regions, with the          
subtraction of the overlapping subregions where both of the variables are \lq\lq         
large\rq\rq, that is          
$| \alpha | > \alpha_0$ and $| \beta | > \beta_0$.  If we decompose the momentum         
transfer         
\be         
\label{eq:a27}         
q_R \; = \; \beta_R p_1 \: + \: \alpha_R p_2 \: + \: q_{R_\perp}         
\ee         
then we have         
\be         
\label{eq:a28}         
s \beta_R \; \simeq \; \mbox{\boldmath $q$}_R^2, \quad s \alpha_R \; \simeq \; - Q^2          
\: - \: \frac{\mbox{\boldmath $\ell$}_-^2}{x_-} \: - \: \frac{\mbox{\boldmath          
$\ell$}_+^2}{x_+}.         
\ee         
Combining this with the fact that all transverse momenta are of order        
$\mu$,  it is easy to show that the          
overlapping regions where $| \alpha |$ and $| \beta |$ are both \lq\lq large\rq\rq~give a          
contribution of relative order $\lapproxeq \mu^2/s \alpha_0 \beta_0$, which vanishes        
in the limit $s \rightarrow \infty$.  Thus $I_{5a}$ is given by the sum of the          
contributions from the three regions.  The result for the central region ($| \alpha | <          
\alpha_0, | \beta | < \beta_0$) is         
\be         
\label{eq:a29}         
I_{5a}^{\rm central} \; = \; \frac{2 \Gamma (1 - \epsilon)}{(4 \pi)^{D/2}} \:          
\frac{\Gamma^2 (\epsilon)}{\Gamma (2 \epsilon)} \: \frac{(\mbox{\boldmath          
$q$}_R^2)^{\epsilon - 1}}{2 s x_-} \: \left [- \ln \left ( \frac{- s \alpha_0          
\beta_0}{\mbox{\boldmath $q$}_R^2} \right ) \: + \: \psi (1) \: - \: \psi (1 -         
\epsilon) \: + \: 2 \psi (\epsilon) \: - \: 2 \psi (2 \epsilon)          
\right ],         
\ee         
where $\psi (x) = \Gamma^\prime (x)/\Gamma (x)$.  We emphasize the              
important fact that the  contribution of the central region is universal   
(i.e. process independent). The above equation gives not only the singular   
$1/\epsilon$ term, but also the general $\varepsilon$ dependence.  For completeness   
we give the contributions from regions $A$ and $B$ which are non-vanishing as   
$\epsilon \rightarrow 0$:   
$$        
I_{5a}^{A} \; = \; \frac{2 \Gamma (1 - \epsilon)}{(4 \pi)^{D/2}} \:          
\frac{(\mbox{\boldmath $q$}_R^2)^{\epsilon - 1}}{2 s x_-} \:  
\left [- \frac{\Gamma^2 (\epsilon)}{\Gamma (2 \epsilon)}\:  
\ln \left ( \frac{x_-}{\beta_0} \right ) \: + \: \frac{1}{\epsilon^2}\:   
+\:\frac{2}{\epsilon}\:\ln \left ( \frac{\Delta_-}  
{\mbox{\boldmath $q$}_R^2} \right )\:+\frac{1}{2}  
\ln ^2 \left ( \frac{\Delta_-}{\mbox{\boldmath $q$}_R^2} \right )\right.  
$$  
\be         
\label{eq:a29A}  
\left.+\:L \left ( 1-\frac{\mbox{\boldmath $q$}_R^2}{\Delta_-}\right )\:  
-\:L \left ( 1-\frac{\Delta_-}{\mbox{\boldmath $q$}_R^2}\right )\:  
+\: \frac{\pi^2}{6}\right ]\:,    
\ee    
where   
$$  
L(x)=\int_0^x\frac{d t}{t} \ln(1-t)\:,   
$$  
and   
\be         
\label{eq:a29B}         
I_{5a}^{B} \; = \; \frac{2 \Gamma (1 - \epsilon)}{(4 \pi)^{D/2}} \:          
\frac{\Gamma^2 (\epsilon)}{\Gamma (2 \epsilon)} \: \frac{(\mbox{\boldmath          
$q$}_R^2)^{\epsilon - 1}}{2 s x_-} \: \left [- \ln \left ( \frac{1}  
{\beta_0} \right ) \: - \: \psi (1) \: +\:  \psi (2 \epsilon)          
\right ].         
\ee         
We checked by an independent          
calculation of the integral of (\ref{eq:a23}), using Feynman parameters, that $I_{5a}$          
is indeed given by the sum of the three contributions, and hence that there is a          
negligible contribution from the regions where both $| \alpha |$ and $| \beta |$ are          
\lq\lq large\rq\rq.         
         
We now study how the basic integral (\ref{eq:a23}) enters the matrix element.  We          
use the same trick that we used before for the metric tensors in the gluon propagators,          
see (\ref{eq:a10}).  In this way we find factors         
\be         
\label{eq:a30}         
\fmslash{p}_2 (- \fmslash{\ell}_- - \fmslash{k}) \fmslash{p}_2 \quad\quad {\rm and}          
\quad\quad \fmslash{p}_1 (\fmslash{p}_2^\prime - \fmslash{k}) \fmslash{p}_1         
\ee         
respectively in the numerator for the upper and lower blocks in Fig.~5a.   
It is clear that we can neglect $\fmslash {k}$ in the central region.   
Since we can omit the contribution from the region where both $| \alpha |$  
and $| \beta |$ are large, we are also able to neglect $\fmslash{k}$ in the numerator of   
the upper (lower) block in          
the $A (B)$ region.  We can then factorize from the numerator the corresponding term          
which arises in the amplitude of Fig.~3a.  To see this we use the anticommutation          
relations of the $\gamma$ matrices to move $\fmslash{\ell}_-          
(\fmslash{p}_2^\prime)$ to act on the spinor $v (\ell_-) (\bar{u} (p_2^\prime))$, and          
then use the Dirac equation.  We also have factorization in colour space.  We require          
negative-signature colour-octet exchange in the $t_R$ channel, and so we need the          
antisymmetric colour-octet projection of the two-gluon state.  The colour structure is          
thus identical to that of the diagrams of Fig.~3.         
         
Therefore it turns out that the contribution of the \lq\lq central\rq\rq~region of Fig.~5a          
to the $\gamma^* q \rightarrow (q \bar{q}) q$ amplitude $T^{(8)}$ of (\ref{eq:a5}) is       
\be        
\label{eq:a31}        
T_{5a}^{\rm central} \; = \; \frac{s}{t_R} \: \langle q|t^i|\bar {q}        
\rangle  \:\Gamma_a \:        
\Gamma_{q^{\prime},q}^{(0)         
i} \: \omega (t_R) \left [ \ln \left ( \frac{-s}{\mbox{\boldmath $q$}_R^  
2} \right ) \; +         
\: \phi (\alpha_0) \: + \: \phi (\beta_0) \right ],        
\ee        
where $\Gamma_a$ is the $\gamma^* \rightarrow q\bar{q}$ Reggeon vertex of         
(\ref{eq:a13}), $\Gamma_{q^{\prime},q}^{(0)         
i}$ is the $q \rightarrow q^{\prime}$ Reggeon vertex of         
(\ref{eq:a11}), $\omega$ is the one-loop contribution to the trajectory function of         
(\ref{eq:a21}), and the $\phi$ functions are        
\be        
\label{eq:a32}        
\phi (z) \; = \; \ln z \: - \: \frac{1}{2} \left [ (\psi (1) \: - \: \psi (1 - \epsilon)         
\right ] \: - \: \psi (\epsilon) \: + \: \psi (2 \epsilon).        
\ee        
Diagrams 5b, 5a$^\prime$ and 5b$^\prime$ can be computed in exactly the same         
way.  The contribution of the central region of Fig.~5b is given by (\ref{eq:a31}) with         
the replacement of $\Gamma_b$ of (\ref{eq:a14}) for $\Gamma_a$; while the         
contributions of 5a$^\prime$ and 5b$^\prime$ are obtained from those of diagrams 5a         
and 5b by the replacement $s \rightarrow -s$ and by changing the overall sign (due to         
antisymmetry in colour space).  Collecting all these four central region contributions         
together we therefore have       
\be        
\label{eq:a33}        
T^{\rm central} \; = \; \frac{2s}{t_R} \: \Gamma_{q\bar{q}, \gamma^*}^{(0) i} \:         
\Gamma_{q^{\prime},q}^{(0) i} \: \omega (t_R) \left [ \frac{1}{2} \ln \left ( \frac{-        
s}{\mbox{\boldmath $q$}_R^2} \right ) \: + \: \frac{1}{2} \ln \left (         
\frac{s}{\mbox{\boldmath $q$}_R^2} \right ) \: + \: \phi (\alpha_0) \: + \:         
\phi (\beta_0) \right ],   
\ee        
where we have used (\ref{eq:a12}).  We see that the logarithmic part of the    
contribution coincides with the first term in the         
expansion of (\ref{eq:a5}) in $\omega (t_R)$; that is, it is responsible for gluon         
Reggeization.  The remaining pieces, $\phi (\alpha_0)$ and $\phi         
(\beta_0)$, must be included as corrections to the corresponding vertices.   
The important fact is that contribution (\ref{eq:a33}) of the central region is {\it         
universal}.  That is this result for $\gamma^* q \rightarrow (q\bar{q}) q$ applies to         
any process $A + a \rightarrow B + b$, as in Fig.~2, with the same expression in        
square brackets, provided that we use the corresponding vertices.        
        
We now turn to the $A$ region ($| \beta | > \beta_0$, see (\ref{eq:a25})).  We first    
consider the combined contribution of the diagrams of Figs.~5a, 5a$^\prime$ and 6a.          
To evaluate diagram 6a we again use the familiar trick (\ref{eq:a10}) for the metric         
tensor in the propagator of the gluon with momentum $q_R$.  In this way we show        
that we can         
factor out the vertex $\Gamma_{q^{\prime},q}^{(0) i}$.  The same factorization        
applies to         
diagrams 5a and 5a$^\prime$ in the $A$ region.  Moreover here we can replace        
\bea        
\label{eq:a34}        
(p_2^\prime - k)^2 \: + \: i0 & = & -s (\beta - \beta_R) (1 - \alpha - \alpha_R) \: - \:         
(\mbox{\boldmath $q$}_R - \mbox{\boldmath $k$})^2 \: + \: i 0 \; \rightarrow \; - s         
\beta \nonumber \\        
& & \\        
(p_2 + k)^2 \: + \: i0 & = & s \beta \: - \: \mbox{\boldmath $k$}^2 \: + \: i 0 \;         
\rightarrow \; s \beta. \nonumber        
\eea        
Then taking into account the difference in signs for diagrams 5a and 5a$^\prime$ due         
to colour antisymmetry, we see that they give equal contributions in the $A$ region.          
After the factorization of the common factor $\Gamma_{q^{\prime},q}^{(0) i}$ we        
find that the remainder does not depend on the properties of the lower block (vertex).         
Thus the sum of the remaining contributions of diagrams 5a, 5a$^\prime$ and 6a must        
give corrections to the vertex $\Gamma_{q\bar{q}, \gamma^*}$.  Moreover this sum        
is given by Fig.~7a, where we have introduced the off-mass-shell        
gluon-gluon-Reggeon vertex of Fig.~8  
\bea        
\label{eq:a35}        
\gamma_{i a b}^{\mu \nu} (k_1, k_2) & = &- \frac{ig}{s} \: T_{ab}^i \biggl [ -        
g^{\mu \nu} \: p_2 \cdot (k_2 - k_1) \: - \: p_2^\mu (2 k_1 + k_2)^\nu \biggr .         
\nonumber \\        
& & \\        
& & + \; p_2^\nu (2 k_2 + k_1)^\mu \: - \: 2 (k_1 + k_2)^2 \: \frac{p_2^\mu         
p_2^\nu}{p_2 \cdot (k_1 - k_2)}  \biggr ] \nonumber        
\eea        
where $T^i$ are the generators in the adjoint representation in colour space;         
$T_{ab}^c = -i f_{cab}$.  
         
In the same way we factor off the lower vertex $\Gamma_{q^\prime, q}^{(0) i}$ from       
the $A$ region contributions of diagrams 5b, 5b$^\prime$ and 6b, and find that their       
sum is given by Fig.~7b.  It is also clear that the $B$ region contributions of diagrams       
5a, b, a$^\prime$, b$^\prime$, in which the upper vertex $\Gamma_{q\bar{q},       
\gamma^*}^{(0) i}$ has been factored off, will give the corrections to the lower       
quark-quark-Reggeon vertex $\Gamma_{q^\prime, q}$.  We are, however, not       
concerned with these latter corrections here.      
      
Now we turn to the remaining two gluon exchange diagrams, 5c, c$^\prime$.  Unlike       
the previous diagrams here both the produced $q$ and $\bar{q}$ interact, and they       
contribute in quite different regions of the Sudakov parameters $\alpha$ and $\beta$.        
Only the region of small $\alpha$, $| \alpha | \sim \mu^2/s$, is important.  The reason   
is that the two virtual fermions in the upper block have momenta components of order       
unity along $p_1$, and $\sim \alpha + O (\mu^2/s)$ along $p_2$.  The denominators   
of their propagators are 
  
\bea  
\label{eq:a36}      
& & \left ( s (x_- + \beta) \left (\alpha + \frac{\mbox{\boldmath $\ell$}_-^2}{sx_-}   
\right ) - (\mbox{\boldmath $\ell$}_- + \mbox{\boldmath $k$})^2  +  i0 \right ) \:  
\times \nonumber \\  
& & \nonumber \\  
& & \times \: \left ( s (x_+ + \beta_R - \beta) \left ( \alpha_R - \alpha +  
\frac{\mbox{\boldmath $\ell$}_+^2}{sx_+} \right )  -  (\mbox{\boldmath $\ell$}_+ +  
\mbox{\boldmath $q$}_R-\mbox{\boldmath $k$})^2  +  i0 \right ),      
\eea      
whereas we find the numerators (after performing the trick with the metric tensors in       
the gluon propagators) do not contain any $\alpha$ dependence.  Therefore the       
integral over $\alpha$ converges with the main contribution coming from the region $|       
\alpha | \sim \mu^2/s$.  Let us now consider the $\beta$ integration.    
The contribution of both diagrams 5c and 5c$^\prime$ to the colour-octet, negative-  
signature amplitude in the $t_R$ channel contains a factor, from the fermion   
propagators of the lower block, of      
\be      
\label{eq:a37}      
\left ( \frac{1}{\beta - \mbox{\boldmath $k$}^2/s + i0} \; + \; \frac{1}{\beta - \beta_R       
+ (\mbox{\boldmath $q$}_R - \mbox{\boldmath $k$})^2/s - i0} \right ).      
\ee      
In the limit $s\rightarrow \infty$ it gives $2P(1/\beta)$, since $\beta_R \simeq   
\mbox{\boldmath $q$}_R^2/s$, where $P$ means the   
principal value. For $| \alpha | \sim \mu^2/s$, the $\beta$ dependence in all the other   
propagators can be neglected at small $\beta$.  Thus the region $|\beta| \leq \beta_0$   
does not contribute.  Hence diagrams 5c, c$^\prime$ contribute only in the $A$   
region.  As       
before, we can combine them with diagram 6c and factorize off the vertex       
$\Gamma_{q^\prime, q}^{(0) i}$.  The remainder can be represented by diagram 6c       
with the off-mass-shell gluon-gluon-Reggeon vertex of (\ref{eq:a35}).      
      
Let us summarize the analysis of the two-gluon exchange diagrams of Fig.~5.  These       
diagrams, which are responsible for gluon Reggeization, also contribute to the       
Reggeon vertices.  The contributions to the $\gamma^* \rightarrow q\bar{q}$ vertex,       
which we are interested in, come both from the central region $(| \alpha | < \alpha_0, |       
\beta | < \beta_0)$ and from the $A$ region $(| \beta | > \beta_0)$.  The contributions       
coming from the $A$ region can be combined with the contributions of Fig.~6 and       
represented by the diagrams of Fig.~7 with the gluon-g  
luon-Reggeon vertex given by       
(\ref{eq:a35}).  The contribution coming from the \lq\lq central\rq\rq~region can be       
extracted from (\ref{eq:a33})      
\be      
\label{eq:a38}      
\Delta \Gamma_{q\bar{q}, \gamma^*}^{i ({\rm central})} \; = \; - \bar{g}^2 \:       
\frac{\Gamma^2 (\epsilon)}{\Gamma (2 \epsilon)} (\mbox{\boldmath       
$q$}_R^2)^\epsilon \: \Gamma_{q\bar{q}, \gamma^*}^{(0) i} \: \phi (\beta_0),      
\ee      
where we have used (\ref{eq:b20}), and where $\phi (z)$ is given by (\ref{eq:a32}).        
The intermediate parameter $\beta_0$ cancels when we combine (\ref{eq:a38}) with     
the contributions of Figs.~7a,b.  (The contribution of diagram 7c does not depend on       
$\beta_0$.)      
      
In addition to the two-gluon exchange contributions, we have also to include    
corrections to the $\gamma^* \rightarrow q\bar{q}$ vertex coming from the diagrams    
of Fig.~9, as well as from the quark self-energy diagrams, and also from half of the    
self-energy contribution of the gluon of momentum $q_R$.  The latter contribution is    
not infrared divergent and will not be discussed further.

{\noindent \bf 3.2~~Loop diagrams within the produced $q\bar{q}$ pair}   
   
The infrared divergent contribution of the quark self-energy diagrams can be found as    
follows.  First the self-energy of the internal fermion    
lines is not infrared divergent.  For the external lines, in dimensional regularization,    
the ultraviolet and infrared divergences cancel.  Therefore it is simplest to determine    
the infrared divergence by taking the ultraviolet divergence with the opposite sign.  It    
gives   
\be   
\label{eq:a39}   
\Delta \Gamma_{q\bar{q}, \gamma^*}^{i (\rm quark~self-energy)} \; = \;    
\Gamma_{q\bar{q}, \gamma^*}^{(0) i} \left ( - \: \frac{\bar{g}^2}{\epsilon} \:    
\frac{C_F}{N_C} \right ).   
\ee   
   
It remains to calculate the divergent contributions of Figs.~7 and 9.  Unfortunately, in    
the Feynman gauge, there are many artificial divergences in the contributions of the    
separate diagrams, which cancel in the sum.  Although calculations can be done more    
easily by the choice of an appropriate gauge, we prefer to work in the Feynman gauge    
in order to escape ambiguities of the light-cone gauges.   
   
Our procedure is to calculate $D$-dimensional integrals over the virtual gluon    
momentum $k$ using the Sudakov components.  We first perform the integration over    
$\alpha$ using the residue theorem.  We then introduce an intermediate parameter    
$x_0$ to consider the collinear singularities in the regions $| \beta | < x_0$ and $| \beta    
| > x_0$ separately.  We take $x_0$ small, $x_0 \ll 1$.  We let it tend to zero but not    
before $\epsilon \rightarrow 0$, so that $\epsilon \ln(1/x_0) \rightarrow 0$ as $\epsilon   
\rightarrow 0$.  On the other hand we take $\beta_0 \rightarrow 0$ before $\epsilon    
\rightarrow 0$.  Recall that all calculations at next-to-leading order are performed by    
first taking $s \rightarrow \infty$ and then $\epsilon \rightarrow 0$.   
   
For $| \beta | < x_0$ all three diagrams of Fig.~7 contribute, as well as diagrams a1    
and b1 of Fig.~9.  At first sight it may seem strange that the diagrams of Fig.~7 are    
divergent in the \lq soft\rq~region.  However recall that they contain the    
gluon-gluon-Reggeon vertex and receive contributions to the $\gamma^* \rightarrow    
q\bar{q}$ vertex from the diagrams of Fig.~5.  The contribution of diagram 7a is   
\be   
\label{eq:a40}   
\Delta \Gamma_{q\bar{q}, \gamma^*}^{(\rm soft~7a)} \; = \; - \bar{g}^2 \left [    
\frac{\Gamma^2 (\epsilon)}{\Gamma (2 \epsilon)} \: (\mbox{\boldmath    
$q$}_R^2)^\epsilon \: \ln \frac{1}{\beta_0} \: - \: \frac{1}{\epsilon^2} \: - \:    
\frac{1}{\epsilon} \: \ln \left ( \frac{\Delta_-}{x_0 x_-} \right ) \right ] \: \Gamma_a,   
\ee   
while that of 7b is obtained by the substitution $q \leftrightarrow \bar{q}$, that is by    
$\Gamma_a \leftrightarrow \Gamma_b, \; \Delta_- \leftrightarrow \Delta_+$ and $x_-    
\leftrightarrow x_+$.  The $\ln (1/\beta_0)$ terms cancel the exactly analogous terms    
coming from the central region, see (\ref{eq:a38}) and (\ref{eq:a32}), as they must.     
Diagram 7c, which is symmetric with respect to the $q \leftrightarrow \bar{q}$    
substitution, is given by   
\be   
\label{eq:a41}   
\Delta \Gamma_{q\bar{q}, \gamma^*}^{(\rm soft~7c)} \; = \; - \bar{g}^2 \left [    
\frac{1}{\epsilon^2} \: + \: \frac{1}{\epsilon} \: \ln \left ( \frac{\Delta_- x_0}{x_-}    
\right ) \right ] \: \Gamma_a \: + \: (q \leftrightarrow \bar{q}).   
\ee   
   
Turning now to diagram 9a1 we have   
\be   
\label{eq:a42}   
\Delta \Gamma_{q\bar{q}, \gamma^*}^{(\rm soft \: 9a1)} \; = \; \bar{g}^2 \:    
\frac{1}{N_C^2} \left [ \frac{1}{\epsilon^2} \: + \: \frac{1}{\epsilon} \: \ln \left (    
\frac{(\ell_+ + \ell_-)^2 x_0^2}{x_+ x_-} \right ) \right ] \Gamma_a,   
\ee   
while the contribution of 9b1 is obtained by the substitution $q \leftrightarrow    
\bar{q}$.  The total contribution from the central region [(\ref{eq:a38}),    
(\ref{eq:a32})] and the soft region [(\ref{eq:a40})--(\ref{eq:a42})], plus the $q    
\leftrightarrow \bar{q}$ diagrams of 7b and 9b1] is   
\be   
\label{eq:a43}   
\Delta \Gamma_{q\bar{q}, \gamma^*}^{i (\rm soft \: + \: central)} \; = \; - \bar{g}^2    
\left    
[ \frac{1}{\epsilon^2} + \frac{1}{\epsilon} \ln (\mbox{\boldmath $q$}_R^2 x_0^2) -    
\frac{1}{N_C^2} \left (\frac{1}{\epsilon^2} + \frac{1}{\epsilon} \ln \left (    
\frac{(\ell_+ + \ell_-)^2 x_0^2}{x_+ x_-} \right ) \right ) \right ] \:    
\Gamma_{q\bar{q}, \gamma^*}^{(0) i}.   
\ee   
   
We now study the diagrams with {\it collinear} singularities.  As mentioned above,    
in the Feynman gauge there are artificial singularities in the separate diagrams, which    
cancel each other in the sum.  The final answer is very simple and could be written    
down straightaway.  Nevertheless we check explicitly that the cancellation indeed    
occurs.   
   
For instance diagram 7a has a singular contribution from the region in which $k$ is    
collinear with $\ell_-$   
\be   
\label{eq:a44}   
\Delta \Gamma_{q\bar{q}, \gamma^*}^{i (\rm 7a \: collinear)} \; = \; - \:    
\frac{\bar{g}^2}{\epsilon} \left ( \ln \frac{x_-}{x_0} \: - \: 1 \right ) \: \langle q | t^i |    
\bar{q} \rangle \Gamma_a \: - \: R_a^i,   
\ee   
with the non-factorizable contribution   
\be   
\label{eq:a45}   
R_a^i \; = \; eg \frac{\bar{g}^2}{\epsilon} \: \int_0^1 \: dz \frac{(1 - z) x_-}{2 (q_R +    
z \ell_-)^2} \: \bar{u} (\ell_+) \: t^i \: \fmslash{\varepsilon}_{\gamma^*} \: v (\ell_-).   
\ee   
It is easy to understand the origin of such non-factorizable terms, and to follow their    
mutual cancellation using the gauge properties of the gluon vertices.  The    
gluon-gluon-Reggeon vertex (\ref{eq:a35}) gives   
\be   
\label{eq:a46}   
\gamma_{i a b}^{\mu \nu} (k_1, k_2) k_2^\nu \; = \; - \: \frac{ig}{s} \: T_{ab}^i \left    
[ k_1^2 p_2^\mu \: + \: (k_2 \cdot p_2) k_1^\mu \right ],   
\ee   
since $p_2 \cdot k_2 \simeq - p_2 \cdot k_1$ as the Reggeon momentum $q_R = k_1    
- k_2$ is almost transverse; thus $p_2 \cdot q_R$ does not grow with $s$, whereas    
$p_2 \cdot k_1$ and $p_2 \cdot k_2$ are proportional to $s$.  A vertex describing the    
interaction of a collinear gluon and a fermion gives us a gluon momentum which has    
to be contracted (via the gluon propagator) with another gluon vertex.  For example,    
in the region where $k$ and $\ell_-$ are collinear, $k \simeq - z \ell_-$, the matrix    
element of diagram 7a contains the factor   
\be   
\label{eq:a47}   
( \fmslash{\ell}_- + \fmslash{k}) \gamma^\nu v(\ell_-) \: \simeq \: (1 - z)    
\fmslash{\ell}_- \gamma^\nu v (\ell_-) \: \simeq \: 2 (1 - z) \ell_-^\nu v (\ell_-) \:    
\simeq \: - 2 \frac{(1 - z)}{z} k^\nu v (\ell_-).   
\ee   
Now we use the gauge properties of the vertices.  For diagram 7a we put $k_2 = k$    
and $k_1 = q_R - k$ into the gluon-gluon-Reggeon vertex of (\ref{eq:a46}).  The first    
term gives a contribution proportional to $\Gamma_a$, and the second term the    
non-factorizable contribution $R_a^i$ of (\ref{eq:a45}), where we recognize the    
gluon propagator with momentum $q_R - k$ with  
 $k = - z \ell_-$.   
   
The contribution of diagram 7b has a form analogous to (\ref{eq:a44}), coming from    
the region where $k$ is collinear with $\ell_+$,   
\be   
\label{eq:a48}   
\Delta \Gamma_{q\bar{q}, \gamma^*}^{i (\rm 7b \: collinear)} \; = \; - \:    
\frac{\bar{g}^2}{\epsilon} \left ( \ln \frac{x_+}{x_0} \: - \: 1 \right ) \: \langle q |    
t^i | \bar{q} \rangle \Gamma_b \: - \: R_b^i,   
\ee   
where   
\be   
\label{eq:a49}   
R_b^i \; \simeq \; - eg \frac{\bar{g}^2}{\epsilon} \: \int_0^1 \: dz \frac{(1 - z)    
x_+}{2 (q_R + z \ell_+)^2} \: \bar{u} (\ell_+) t^i \:    
\fmslash{\varepsilon}_{\gamma^*} v (\ell_-).   
\ee   
   
Diagram 7c has contributions both from the region where $k$ is collinear with    
$\ell_+$ and from the region where $q_R - k$ is collinear with $\ell_-$.  It has the    
non-factorizable form   
\be   
\label{eq:50}   
\Delta \Gamma_{q\bar{q}, \gamma^*}^{i (\rm 7c \: collinear)} \; = \; R_a^i \: + \:    
R_b^i \: + \: R_-^i \: + \: R_+^i,   
\ee   
where $R_a^i$ and   
\be   
\label{eq:a51}   
R_-^i \; = \; - eg \frac{\bar{g}^2}{\epsilon} \: \int_{x_0/x_-}^1 \: dz \frac{(1 -    
z)}{z} \: \bar{u} (\ell_+) t^i \; \frac{\fmslash{p}_2 (\fmslash{\ell}_+ +    
\fmslash{q}_R    
+ z \fmslash{\ell}_-) \fmslash{\varepsilon}_{\gamma^*}}{s (\ell_+ + q_R + z \ell_-   
)^2} \: v (\ell_-)   
\ee   
come from the region where $q_R - k$ is collinear with $\ell_-$.  On the other hand    
$R_b^i$ and   
\be   
\label{eq:a52}   
R_+^i \; = \; eg \frac{\bar{g}^2}{\epsilon} \: \int_{x_0/x_+}^1 \: dz \frac{(1 -    
z)}{z} \: \bar{u} (\ell_+) t^i \; \frac{\fmslash{\varepsilon}_{\gamma^*}    
(\fmslash{\ell}_- + \fmslash{q}_R + z \fmslash{\ell}_+) \fmslash{p}_2}{s (\ell_- +    
q_R + z \ell_+)^2} \: v (\ell_-)   
\ee   
come from the region where $k$ is collinear with $\ell_+$.  We see that the    
contributions $R_a^i$ and $R_b^i$ cancel with the corresponding contributions of    
diagrams 7a and 7b respectively.   
   
Contributions $R_\pm^i$ cancel with analogous contributions from diagrams 9a1,    
a2 and 9b1, b2 respectively, which have the form   
\bea   
\label{eq:a53}   
\Delta \Gamma_{q\bar{q}, \gamma^*}^{i (\rm 9a1 \: collinear)} & = & -    
\frac{1}{N_C^2} R_+^i \: - \: R_1^i  \\   
& & \nonumber \\   
\label{eq:a54}   
\Delta \Gamma_{q\bar{q}, \gamma^*}^{i (\rm 9a2)} & = & - \:    
\frac{\bar{g}^2}{\epsilon} \: \frac{2 C_F}{N_C} \left ( \ln \frac{x_+}{x_0} \: - \: 1    
\right ) \: \langle q | t^i | \bar{q} \rangle \Gamma_a \: - \: \frac{2 C_F}{N_C} \: R_+^i    
\\   
& & \nonumber \\   
\label{eq:a55}   
\Delta \Gamma_{q\bar{q}, \gamma^*}^{i (\rm 9b1 \: collinear)} & = & - \:    
\frac{1}{N_C^2} \: R_-^i \: - \: R_2^i \\   
& & \nonumber \\   
\label{eq:a56}   
\Delta \Gamma_{q\bar{q}, \gamma^*}^{i (\rm 9b2)} & = & - \:    
\frac{\bar{g}^2}{\epsilon} \: \frac{2 C_F}{N_C} \left ( \ln \frac{x_-}{x_0} \: - \: 1    
\right ) \: \langle q | t^i | \bar{q} \rangle \Gamma_b \: - \: \frac{2 C_F}{N_C} \: R_-^i,   
\eea   
where we have not used the \lq collinear\rq~superscript for those diagrams which do    
not have soft singularities.  These contributions contain further artificial    
non-factorizable singularities   
\bea   
\label{eq:a57}   
R_1^i & = & eg \frac{\bar{g}^2}{\epsilon} \: \frac{1}{N_C^2} \: \int_{x_0/x_-}^1 \:    
dz  \frac{(1 - z)}{z} \: \bar{u} (\ell_+) t^i \; \frac{\fmslash{\varepsilon}_{\gamma^*}    
(\fmslash{q}_R + (1 - z) \fmslash{\ell}_-) \fmslash{p}_2}{(q_R + (1 - z) \ell_-)^2} \:    
v (\ell_-) \\   
& & \nonumber \\   
\label{eq:a58}   
R_2^i & = & - eg \frac{\bar{g}^2}{\epsilon} \: \frac{1}{N_C^2} \: \int_{x_0/x_+}^1    
\: dz \frac{(1 - z)}{z} \: \bar{u} (\ell_+) t^i \; \frac{\fmslash{p}_2 (\fmslash{q}_R +    
(1 -    
z) \fmslash{\ell}_+) \fmslash{\varepsilon}_{\gamma^*}}{(q_R + (1 - z) \ell_+)^2} \:    
v (\ell_-).   
\eea   
They in turn cancel with the contributions of diagrams 9a3, b3.   
\bea   
\Delta \Gamma_{q\bar{q}, \gamma^*}^{i (\rm 9a3)} & = &    
\frac{\bar{g}^2}{N_C^2 \epsilon} \left ( \ln \frac{x_-}{x_0} \: - \: 1 \right ) \: \langle    
q | t^i | \bar{q} \rangle \Gamma_a \: + \: R_1^i \nonumber \\   
& & \nonumber \\   
\Delta \Gamma_{q\bar{q}, \gamma^*}^{i (\rm 9b3)} & = &    
\frac{\bar{g}^2}{N_C^2 \epsilon} \left ( \ln \frac{x_+}{x_0} \: - \: 1 \right ) \: \langle    
q | t^i | \bar{q} \rangle \Gamma_b \: + \: R_2^i. \nonumber   
\eea   
   
Taking the sum of all these contributions and adding to it the quark self-energy    
contribution (\ref{eq:a39}), which we may call collinear as well, we obtain   
\be   
\label{eq:a59}   
\Delta \Gamma_{q\bar{q}, \gamma^*}^{i (\rm collinear)} \; = \; - \: \frac{2    
\bar{g}^2}{\epsilon} \: \frac{C_F}{N_C} \left ( \ln \frac{x_+ x_-}{x_0^2} \: - \:    
\frac{3}{2} \right ) \Gamma_{q\bar{q}, \gamma^*}^{(0) i}.   
\ee   
As mentioned earlier, this answer could have been written \lq by hand\rq, since the    
expression in brackets is easily recognized as arising from the quark splitting function   
$$   
\left ( \int_{x_0/x_+}^1 \: + \: \int_{x_0/x_-}^1 \right ) \: dz \: P_{gq} (z) \; = \; \left (    
\int_{x_0/x_+}^1 \: + \: \int_{x_0/x_-}^1 \right ) \: dz \: \frac{1}{2z} \left (1 \: + \: (1    
- z)^2 \right ).   
$$   
We see that the verification of (\ref{eq:a59}) in the Feynman gauge is complicated by    
the presence of spurious singular terms which mutually cancel.  Nevertheless we    
prefer to use this gauge so as to minimize the possible error in the subsequent    
evaluation of the non-singular terms.   
   
\medskip   
\noindent {\bf 3.3~~Total singular virtual contribution}   
   
The total singular part of the virtual correction to the Reggeon vertex for $\gamma^*    
\rightarrow q\bar{q}$ is given by the sum of the \lq soft + central\rq~contribution of    
(\ref{eq:a43}) and the \lq collinear\rq~contribution of (\ref{eq:a59}).  It is   
\be   
\label{eq:a60}   
\Delta \Gamma_{q\bar{q}, \gamma^*}^{i} \; = \; - \: \bar{g}^2 \left    
[\frac{1}{\epsilon^2} \: + \: \frac{1}{\epsilon} \left ( \ln (x_+ x_- \mbox{\boldmath    
$q$}_R^2) \: - \: \frac{3}{2} \right ) \: - \: \frac{1}{N_C^2} \left    
(\frac{1}{\epsilon^2} \: + \: \frac{1}{\epsilon} \left \{ \ln (\ell_+ + \ell_-)^2 \: - \:    
\frac{3}{2} \right \} \right ) \right ] \Gamma_{q\bar{q}, \gamma^*}^{(0) i}.   
\ee   
   
\bigskip    
   
\noindent {\large \bf 4.~~Real gluon emission}      
     
Now we consider the contribution to the photon impact factor from the     
quark-antiquark-gluon intermediate state. First, we have to find the      
$\gamma ^{*}R\rightarrow q\bar qg$ effective vertex. In     
principle, we could use the same approach as for the       
$\gamma^{*}R\rightarrow q\bar q$  vertex. Namely, we could calculate the amplitude     
for the conversion of a virtual photon into a $q\bar{q}g$ state in the high     
energy collision of the photon with a quark, take the part of this amplitude     
with a colour-octet state in the $t$ channel and determine the      
$\gamma ^{*}R\rightarrow q\bar qg$ vertex by comparison of this part with     
the Regge form (\ref{eq:a5}). Since we need the vertex only in the Born     
approximation, it is sufficient to calculate the conversion amplitude in the same     
approximation and to take the Regge form in the lowest approximation in the     
coupling constant.     
     
The expression for the vertex can be obtained immediately, if we represent it     
by the diagrams of Fig.~10 with the vertex factor $igt^i\fmslash {p}_2/s$ for     
the interaction of the Reggeon with quarks, and by the vertex factor    
$\gamma _{iab}^{\mu \nu }$ (\ref{eq:a35}) for the interaction of the Reggeon with    
gluons. It is straightforward to check that this direct calculation gives exactly the same      
$\gamma ^{*}R\rightarrow q\bar qg$ vertex as the method described      
above.      
     
In the following we denote the momentum, polarisation vector and colour index of the    
emitted gluon by $k, e$ and $c$ respectively.  The Sudakov decomposition is      
\begin{equation}     
k\;=\;\beta p_1+\frac{\mbox{\boldmath $k$}^2}{s\beta }p_2\:+\:k_{\perp },     
\quad\quad \mbox{\boldmath $k$}^2\;\equiv -k_{\perp }^2.  \label{eq:a61}     
\end{equation}     
The vertex $\gamma ^{*}R\rightarrow q\bar qg $ has the following  properties.     
\begin{itemize}     
\item[1.] In the \lq soft\rq~region     
\begin{equation}     
\beta \ll 1, \quad\quad \mbox{\boldmath $k$}^2\ll \mu ^2,  \label{eq:a62}     
\end{equation}     
the vertex is given by the formula for accompanying soft bremsstrahlung,     
i.e.     
\begin{equation}     
\Gamma _{q\bar qg,\gamma ^{*}}^i(q_R)=\Gamma _{q\bar q,\gamma     
^{*}}^{}(q_R)g\left[ \left( \frac{\ell_{+}}{k . \ell_{+}}-\frac{p_2}{k . p_2}\right)     
^\mu \langle q|t^ct^i|\bar q \rangle -\left( \frac{\ell_{-}}{k . \ell_{-}}-\frac{p_2}{k .    
p_2}\right)^\mu \langle q|t^it^c|\bar q \rangle \right] e_\mu ^{*}.  \label{eq:a63}     
\end{equation}     
     
\item[2.] In the \lq central\rq~region     
\begin{equation}     
\beta \ll 1,\quad\quad \mbox{\boldmath $k$}^2\gg \beta \mu ^2,  \label{eq:a64}     
\end{equation}     
the vertex factorizes into a product of two Reggeon vertices, i.e     
\begin{equation}     
\Gamma _{q\bar qg,\gamma ^{*}}^i(q_R)=\Gamma _{q\bar q,\gamma     
^{*}}^j(q_{R^{^{\:\prime }}}) \left ( \frac{-1}{\mbox{\boldmath    
$q$}_{R^{\prime}}^2}\right ) \gamma _{ji}^c(q_{R^{^{\:\prime }}},\:q_R),     
\label{eq:a65}     
\end{equation}     
where  $q_{R^{^{\:\prime }}}\:=\:q_R+k$, and where    
$\gamma_{ji}^c(q_{R^{^{\:\prime }}},\:q_R)$, the effective vertex for gluon    
production in a    
Reggeon-Reggeon collision \cite{LLA}, is given by   
\begin{equation}     
\gamma _{ji}^c(q_{R^{^{\:\prime }}},\:q_R)=gT_{ji}^c\left[ -q_{R^{\prime}     
\bot}-q_{R\bot }+(\mbox{\boldmath $k$}^2-2\mbox{\boldmath $q$}_{R^{\prime}}     
^2)\frac{\beta p_1}{\mbox{\boldmath $k$}^2}-(\mbox{\boldmath $k$}^2-     
2\mbox{\boldmath $q$}_R^2)\frac{p_2}{s\beta }\right] ^\mu e_\mu ^{*}.      
\label{eq:a66}     
\end{equation}    
\end{itemize}    
Note that in (\ref{eq:a63}) the gluon momentum is neglected in     
the vertex $\Gamma _{q\bar q,\gamma ^{*}}$ . The neglect is valid in the whole    
region (\ref{eq:a62}),      
not only when $2(k . \ell_{\pm })\ll \mu ^2,$ (i.e.~when $\mbox{\boldmath $k$}^2\ll    
\beta \mu^2$) where it is obviously true. In fact, it is a consequence of Gribov's     
theorem \cite{GR} concerning the region of applicability of formulas for     
accompanying soft bremsstrahlung.  Thus the factorization (\ref{eq:a63}) in     
the region (\ref{eq:a62}) has a general nature.  The same can be said about    
(\ref{eq:a65}). This equation was also checked by direct calculation; but, in fact,  it    
also has quite a general nature and is connected with gluon Reggeization in QCD.      
     
The important point is that the soft (\ref{eq:a62}) and central (\ref{eq:a64})     
regions overlap and cover the entire region $\beta <x_0$ with $x_0 \ll 1$.  Let us    
denote      
\begin{equation}     
\frac 1{\sqrt{N_C^2-1}}\sum_{\{q\bar q g\}}\:     
\Gamma _{q\bar q g,\gamma^{*}}^{(0)i}(q_R)\:     
\Gamma _{q\bar q g,\gamma ^{*\:^{\prime }}}^{(0)i}(q_R^{\prime})^*\:     
=\:f_{q\bar q\:g},   
\label{eq:a67}     
\end{equation}     
c.f.~(\ref{eq:a16a}).  The contribution of the $q\bar q\:g$ intermediate state to the    
impact factor is given by the integral of the function $f_{q\bar q\:g}$ over      
\[     
\frac{dM_{q\bar q\:g}^2}{2\pi }\:d\rho _{q\bar q\:g}\;=2\delta     
(1-x_{+}-x_{-}-\beta)(2\pi )^{D-1}\delta (\mbox{\boldmath $\ell$}_{+}+%
\mbox{\boldmath$\ell$}_{-}+\mbox{\boldmath $q$}_R+\mbox{\boldmath $k$})\;      
\]     
\begin{equation}     
\times \frac{dx_{+}}{2x_{+}}\:\frac{dx_{-}}{2x_{-}}\frac{d\beta}{2\beta}     
\frac{d^{D-2}\ell _{+}}{(2\pi )^{D-1}}\frac{d^{D-2}\ell _{-}}{(2\pi )^{D-1}}     
\frac{d^{D-2}k}{(2\pi )^{D-1}},  \label{eq:a68}     
\end{equation}      
c.f.~(\ref{eq:a17}).  In the soft region we find, using (\ref{eq:a63}), that      
\[     
f_{q\bar q\:g}=f_{q\bar q}(x_{\pm }, \mbox{\boldmath $\ell$}_{\pm };     
\mbox{\boldmath $q$}_R, \mbox{\boldmath $q$}_R^{\prime})  \: \times   
\]     
\begin{equation}     
\times 2 g^2 N_C \left[ \frac{x_+^2}{(x_+\mbox{\boldmath $k$}-     
\beta\mbox{\boldmath $\ell$}_+)^2}+\frac{x_-^2}{(x_-\mbox{\boldmath $k$}-     
\beta\mbox{\boldmath $\ell$}_-)^2}-\frac{1}{N_C^2}\frac{\beta^2     
(x_+\mbox{\boldmath $\ell$}_- -x_-\mbox{\boldmath $\ell$}_+)^2}     
{(x_+\mbox{\boldmath $k$}-\beta\mbox{\boldmath $\ell$}_+)^2     
(x_-\mbox{\boldmath $k$}- \beta\mbox{\boldmath $\ell$}_-)^2}\right],      
\label{eq:a69}     
\end{equation}     
where $f_{q\bar q}$ is defined by (\ref{eq:a16a}).   On the other hand in the central    
region it follows from (\ref{eq:a65}), (\ref{eq:a66}) that      
\begin{equation}     
f_{q\bar q\:g}=f_{q\bar q}(x_{\pm }, \mbox{\boldmath $\ell$}_{\pm };      
\mbox{\boldmath $q$}_{R^{\prime}},    
\mbox{\boldmath $q$}_{R^{\prime}}^{\prime})\:\frac{2(2\pi)^{D-1}}     
{\mbox{\boldmath $q$}_{R^{\prime}}^2     
\mbox{\boldmath $q$}_{R^{\prime}}^{\prime\:2}}\:K_r^{(0)}(\mbox{\boldmath     
$q$}_{R^{\prime}},     
\mbox{\boldmath $q$}_{R}; \mbox{\boldmath $q$}),    
\label{eq:a70}     
\end{equation}     
where $\mbox{\boldmath $q$}_{R^{\prime}}\:=\mbox{\boldmath $q$}_R \: +     
\mbox{\boldmath $k$}; \,\,\,\, \mbox{\boldmath $q$}_{R^{\prime}}^{\prime}     
\:=\: \mbox{\boldmath $q$}_{R^{\prime}} \:-\:\mbox{\boldmath $q$}$ and      
$K_r^{(0)}$ is the part of the non-forward LLA kernel entering      
(\ref{eq:a19}) and defined by (\ref{eq:a20}).   Comparing (\ref{eq:a69}) and    
(\ref{eq:a70}) one discovers an interesting result.  The region of the applicability of    
the  central form (\ref{eq:a70}) for $f_{q\bar{q}g}$ is larger than (\ref{eq:a64}) and,    
in fact, is given by      
\begin{equation}     
\beta \ll 1,\quad\quad \mbox{\boldmath $k$}^2\gg \beta^2 \mu ^2.  \label{eq:a71}     
\end{equation}         
     
It is convenient to consider the contribution of the central region      
together with the subtraction term in (\ref{eq:a19}). This term has      
two types of infrared singularities: the singularities contained in the      
trajectory function $\omega(t)$, and the singularities coming from the integration      
over $\mbox{\boldmath $q$}_{R^{\prime}}$.  Owing to properties   
(\ref{eq:a17b}) of the impact factor, the latter singularities are only due to    
the singular behaviour of the real part of $K_r^{(0)}(\mbox{\boldmath    
$q$}_{R^{\prime}}, \mbox{\boldmath $q$}_{R}; \mbox{\boldmath $q$})\:$ at      
$\mbox{\boldmath $q$}_{R^{\prime}}\:-\:\mbox{\boldmath $q$}_{R} \:=\:     
\mbox{\boldmath $k$}\:=\:0$.  It is useful to note that in (\ref{eq:a19}) $\ln (s_0)$    
appears as a factor of the total non-forward BFKL kernel,  
\begin{equation}     
K(\mbox{\boldmath $q$}_{R^{\prime}},  
\mbox{\boldmath $q$}_{R}; \mbox{\boldmath $q$})\: = \:      
\mbox{\boldmath $q$}_R^2\:\mbox{\boldmath $q$}_{R}^{\prime \: 2}\:     
\delta(\mbox{\boldmath $q$}_{R^{\prime}}\:-\:      
\mbox{\boldmath $q$}_R)\:(\omega(t_R)\: +\:\omega(t_R^{\prime}))\:     
+\:K_r (\mbox{\boldmath $q$}_{R^{\prime}},   
\mbox{\boldmath $q$}_{R}; \mbox{\boldmath $q$}),      
\label{eq:a72}     
\end{equation}      
taken in the Born approximation. Since the total kernel is infrared safe,      
the infrared singularities do not depend on the value of $s_0$, so that      
we can take any suitable choice of $s_0$ to analyse them. It is convenient to take      
$s_0$ sufficiently large so that $s_0 \gg \mu ^2$, but such that at the same time it    
satisfies $s_0 \ll  \mu ^2/x_0^2$.     
      
Then we can evaluate the integral over $f_{q\bar{q}g}$ in the region $\beta \leq x_0$    
as the sum of      
two subregions, in the first (second) of which   $\mbox{\boldmath $k$}^2$      
is less (larger) than $\beta^2s_0$. The contribution of the first region      
can be calculated with the help of (\ref{eq:a69}).  Its singular part is     
\begin{equation}     
\int d\Phi _{\gamma^{*\:\prime },\gamma^*}^{(0)}     
(\mbox{\boldmath $q$}_R,\mbox{\boldmath $q$})  \:   
2\bar g ^2\left[\frac{1}{\epsilon^2}+\frac{1}{\epsilon}\ln(x_0^2s_0)-     
\frac{1}{N_C^2}\left \{ \frac{1}{\epsilon^2}+\frac{1}{\epsilon}     
\ln \left (\frac{x_0^2(\ell_-+\ell_+)^2}{x_+x_-}\right) \right \} \right],      
\label{eq:a73}     
\end{equation}      
where  
\begin{equation}     
d\Phi_{\gamma^{*\:\prime},\gamma^*}^{(0)}     
(\mbox{\boldmath $q$}_R,\mbox{\boldmath $q$})=     
f_{q\bar q}(x_{\pm }, \mbox{\boldmath $\ell$}_{\pm };   
\mbox{\boldmath $q$}_{R},   
\mbox{\boldmath $q$}_{R}^  
{\prime})     
\frac{dM_{q\bar q}^2}{2\pi }\:d\rho _{q\bar q},      
\label{eq:a74}     
\end{equation}     
with the function $f_{q\bar q}$ defined in (\ref{eq:a16a}) and the phase      
space element in (\ref{eq:a17}), so that      
\begin{equation}     
\int d\Phi _{\gamma^{*\:\prime },\gamma^*}^{(0)}     
(\mbox{\boldmath $q$}_R,\mbox{\boldmath $q$})=     
\Phi _{\gamma^{*\:\prime },\gamma^*}^{(0)}     
(\mbox{\boldmath $q$}_R,\mbox{\boldmath $q$}),   
\label{eq:a75}     
\end{equation}     
see (\ref{eq:a7}) and (\ref{eq:a16a}).  The contribution of the second region can be    
calculated with the help of      
(\ref{eq:a70}). As already mentioned, it is convenient to consider this      
contribution  together with the subtraction term in (\ref{eq:a19}).       
Our choice of $s_0$ makes such a consideration especially convenient because      
of the total cancellation  of the singular part of this contribution with the part of the      
subtraction term involving $K_r^{(0)}$. So, the singular contribution from      
the second region and the subtraction term is simply   
\begin{equation}     
\Phi _{\gamma^{*\:\prime },\gamma^*}^{(0)}     
(\mbox{\boldmath $q$}_R,\mbox{\boldmath $q$})     
\frac{\bar g ^2}{\epsilon}\ln\left(\frac{\mbox{\boldmath $q$}_R^2     
\mbox{\boldmath $q$}_R^{\prime\:2}}{s_0^2}\right).     
\label{eq:a76}        
\end{equation}     
The dependence on $s_0$ vanishes in the sum of (\ref{eq:a73}) and      
(\ref{eq:a76}), as it must.      
     
It remains to consider the singular contributions from the region      
$\beta \geq x_0$. These contributions have a collinear nature.      
For real emission we can use, without uncertainty, a physical gauge. In such a gauge,    
the $\gamma^* R \rightarrow q\bar q g$ vertex in the region of quasi-collinearity      
of the gluon and the quark (antiquark) momenta is given by the diagrams      
with a gluon line attached to the external quark (antiquark) line, Fig.~10a1, b1    
(Fig.~10a3, b3).  They can be easily calculated by      
the quasi-real electron method  \cite{BFK}.  In this way the singular contribution      
from the region $\beta \geq x_0$ is found to be    
\[     
\int d\Phi _{\gamma^{*\:\prime },\gamma^*}^{(0)}     
(\mbox{\boldmath $q$}_R,\mbox{\boldmath $q$})\,     
\frac{\bar g ^2}{\epsilon}\frac{2C_F}{N_C}\left[ \left (\int _{x_0/x_+}^1     
+ \int _{x_0/x_-}^1 \right ) \: \frac{dz}{z}(1+(1-z)^2) \right]      
\]     
\begin{equation}     
= \int d\Phi _{\gamma^{*\:\prime },\gamma^*}^{(0)}     
(\mbox{\boldmath $q$}_R,\mbox{\boldmath $q$})\,     
\frac{\bar g ^2}{\epsilon}\frac{2C_F}{N_C}\left[2\ln \left (\frac{x_+x_-}{x_0^2}    
\right ) - 3 \right].      
\label{eq:a77}        
\end{equation}     
   
This completes the study of the real gluon emission contribution to      
the photon impact factor. The singular parts of this contribution together      
with the subtraction term in eq. (\ref{eq:a19}) are given      
by (\ref{eq:a73}), (\ref{eq:a76}) and  (\ref{eq:a77}).   \\   
        
\noindent {\large {\bf 5.~~Cancellation of the infrared singularities}}     
     
Let us denote the singular part of the virtual photon impact factor as      
the sum       
\begin{equation}     
\Phi _{\gamma^{*\:\prime },\gamma^*}     
(\mbox{\boldmath $q$}_R,\mbox{\boldmath $q$})_{\rm{singular}} =     
\int d\Phi _{\gamma^{*\:\prime },\gamma^*}^{(0)}     
(\mbox{\boldmath $q$}_R,\mbox{\boldmath $q$})     
\left[\delta_{\rm virtual}^{\rm soft \: + \: central}+\delta_{\rm virtual}^{\rm    
collinear}+  \delta_{\rm real}^{\rm soft}+\delta_{\rm real}^{\rm central}+    
\delta_{\rm real}^{\rm collinear}     
\right].      
\label{eq:a78}        
\end{equation}     
Here the first two terms in the square brackets are the singular parts      
connected with the virtual corrections to the      
$\gamma^* R\rightarrow q\bar q$ vertex and      
coming from the central and soft regions (\ref{eq:a43}) and the      
collinear regions (\ref{eq:a59})   
\begin{eqnarray}     
\delta_{\rm virtual}^{\rm soft \: + \: central} & = & -\bar g    
^2\left[\frac{2}{\epsilon^2}+     
\frac{1}{\epsilon}\ln(x_0^4\mbox{\boldmath $q$}_R^2     
\mbox{\boldmath $q$}_R^{\prime\:2})-     
\frac{2}{N_C^2}\left(\frac{1}{\epsilon^2}+\frac{1}{\epsilon}     
\: \ln \left (\frac{x_0^2(\ell_-+\ell_+)^2}{x_+x_-} \right )\right)\right].  \label{eq:a79}      
\\   
& & \nonumber \\     
\delta_{\rm virtual}^{\rm collinear} & = & -\frac{\bar    
g^2}{\epsilon}\frac{2C_F}{N_C}     
\left[2 \ln \left (\frac{x_+x_-}{x_0^2}\right ) - 3 \right].      
\label{eq:a80}        
\end{eqnarray}     
The next three terms are the      
singular parts of the contribution of the real gluon emission taken      
together with the subtraction term in (\ref{eq:a19}).  They are given by    
(\ref{eq:a73}), (\ref{eq:a76}) and  (\ref{eq:a77})   
\bea   
\delta_{\rm real}^{\rm soft} & = & 2\bar g ^2\left[\frac{1}{\epsilon^2}+     
\frac{1}{\epsilon}\ln(x_0^2s_0) -    
\frac{1}{N_C^2}\left(\frac{1}{\epsilon^2}+\frac{1}{\epsilon}     
\ln \left (\frac{x_0^2(\ell_-+\ell_+)^2}{x_+x_-}\right ) \right)\right],  \label{eq:a81}     
\\   
& & \nonumber \\      
\delta_{\rm real}^{\rm central} & = & \frac{\bar g ^2}{\epsilon}\ln\left(     
\frac{\mbox{\boldmath $q$}_R^2\mbox{\boldmath $q$}_R^{\prime\:2}}{s_0^2}     
\right),      
\label{eq:a82}        
\eea     
and     
\begin{equation}     
\delta_{\rm real}^{\rm collinear} = \frac{\bar g ^2}{\epsilon}\frac{2C_F}{N_c}     
\left[2\ln \left (\frac{x_+x_-}{x_0^2}\right ) - 3 \right].      
\label{eq:a83}      
\end{equation}     
We see that the sum of the contributions (\ref{eq:a79})-(\ref{eq:a83}) is zero.  That is    
the impact factor is infrared safe.   \\   
     
\noindent {\large {\bf 6.~~Infrared safety of impact factors in the general case}}     
      
In the preceding Sections we have explicitly calculated the infrared contributions to    
the impact factor of the virtual photon coming from various sources, and have    
demonstrated their cancellation.      
In this Section we present general arguments, applicable  to any      
impact factor, to prove its infrared safety. These arguments are based      
on well known  theorems \cite{KLN} about the absence of infrared      
divergences in totally inclusive quantities. Though these theorems are      
commonly used for cross sections and for the imaginary parts of forward      
scattering amplitudes, they apply equally well to non-forward scattering.      
     
Therefore we could be sure of the infrared safety of the impact factors provided that    
Reggeons were a kind of off-mass-shell partons, described by some fields, and      
the impact factors were the imaginary parts of the amplitudes for the scattering of      
these \lq partons\rq~off particles. But this is not the case for several      
reasons. First, the impact factors contain       
the subtraction (see (\ref{eq:a19}) where we have singularities due to the gluon    
trajectory function $\omega (t)$ as well as arising from $K_r^{(0)}$). Second, the    
region of      
integration over the phase space volume in (\ref{eq:a19}) is restricted by      
a theta-function. And finally, the Reggeon-particle vertices $\Gamma^i$      
in (\ref{eq:a19}) are not true field theory amplitudes.      
     
The first two reasons are evident and do not need further explanation. Let us amplify      
the third reason.  We noted that at the Born level the vertices      
$\gamma^*R \rightarrow q\bar q$ and $\gamma^*R \rightarrow q\bar q g$      
are given by Feynman diagrams. It means, that these vertices      
appear as  amplitudes of the effective field theory \cite{A10}.      
An analogous statement is applicable in the general case, but it is only valid in the      
Born approximation.  Recall that the corrections to the      
$\gamma^*R \rightarrow q\bar q$ vertex consist of two parts. One contribution, which    
comes from the central region of virtual gluon momenta (\ref{eq:a25}),      
is given by (\ref{eq:a33}) and is universal, i.e. process independent.      
The other contribution, which comes from the $A$-region, is given by Feynman    
diagrams of the effective field theory with      
the restriction $|\beta| \geq \beta_0$ on the Sudakov parameters of the gluon momenta.       
This last statement again has a general natur  
e, i.e.~in the general case the one-loop    
corrections to Reggeon-particle      
vertices are given by the sum of the universal contribution (\ref{eq:a33}) and a part      
given by the effective field theory with the restriction $|\beta| \geq \beta_0$ on gluon    
momenta.  We can apply arguments  \cite{KLN} about the      
absence of infrared      
divergences to the imaginary parts of Reggeon-particle scattering amplitudes      
in this theory. From the above discussion it is clear how impact      
factors differ from such imaginary parts.  They differ by   
\begin{itemize}     
\item[1)] the existence of the subtraction term;      
\item[2)] the presence of the contribution of (virtual and real) gluons with      
$|\beta| \leq \beta_0$;      
\item[3)] the limitation of the integration region over the phase space of the      
produced particles.      
\end{itemize}   
So, to be convinced that the impact factors are indeed infrared safe, we need  to check    
the cancellation of the infrared contributions to the impact factor coming from these    
three sources.      
      
It is convenient to use a notation analogous to (\ref{eq:a78}) for the singular part of    
the next-to-leading correction to the impact factor   
\be   
\Phi_{A^\prime A} (\mbox{\boldmath $q$}_R, \mbox{\boldmath $q$})_{\rm    
singular} \; = \; \int \: d \Phi_{A^\prime A}^{(0)} (\mbox{\boldmath $q$}_R,    
\mbox{\boldmath $q$}) \: \left [\delta_1 + \delta_2^{\rm virtual} + \delta_2^{\rm    
real} + \delta_3 \right ].   
\label{eq:b83}   
\ee   
Then, taking into account the properties (\ref{eq:a17b}) of the impact factors for      
colourless objects, we obtain from (\ref{eq:a19})-(\ref{eq:a21})     
$$      
\delta_1\: =\:-\;\frac 12 \left[\:\int \:\frac{d^{D-2}k}     
{\mbox{\boldmath $k$}^2}\frac{2g^2N_C}{(2\pi)^{D-1}}     
\:\ln \left( \frac{s_\Lambda ^2}{\mbox{\boldmath $k$}^2s_0}\right)     
+\omega(t_R)\:\ln \left( \frac{\mbox{\boldmath $q$}_R^2}     
{s_0}\right) \:+\:\omega (t_R^{\prime })\:\ln \left( \frac{%
\mbox{\boldmath $q$}_R^{\prime 2}}{s_0}\right)  \right]_{\rm singular}     
$$     
\begin{equation}     
=\:-\bar g ^2\left(\frac{2}{\epsilon^2} +\frac{1}{\epsilon}\ln\left(     
\frac{s_{\Lambda}^4}{\mbox{\boldmath $q$}_R^{\prime 2}     
\mbox{\boldmath $q$}_R^2}\right)\right).     
\label{eq:a84}     
\end{equation}      
Recall from Section 3 that the contribution of  virtual gluons with      
$|\beta| \leq \beta_0$ is universal.  It gives the correction   
(\ref{eq:a38}) to Born vertex.      
Therefore, we obtain      
\begin{equation}     
\delta_2^{\rm virtual} =- \bar g ^2\left(\frac{2}{\epsilon^2}      
+\frac{4}{\epsilon}\ln {\beta_0}     
+\frac{1}{\epsilon}\ln\left(\mbox{\boldmath $q$}_R^{\prime 2}     
\mbox{\boldmath $q$}_R^2\right)\right).      
\label{eq:a85}     
\end{equation}     
It is appropriate to consider the contribution of real      
gluons with $|\beta| \leq \beta_0$ together with the      
limitation $M_a^2 \leq s_{\Lambda}$ of the integration region over the phase space    
of the produced partons in (\ref{eq:a19}).  This limitation does not play any role in the    
LLA, and becomes important only for gluon emission.  Thus, in fact it is a limitation    
on the phase space of the emitted gluon to the region   
$$     
\frac{\mbox{\boldmath $k$}^2}{\beta} \leq s_{\Lambda}.     
$$        
Recall that we take limit $\beta_0 \rightarrow 0$ before $\epsilon      
\rightarrow 0$, so that for the calculation of the contribution of real      
gluons with $|\beta| \leq \beta_0$ we can use the multi-Regge      
factorisation formula analogous to (\ref{eq:a70}).      
Therefore we obtain      
\bea   
\delta_2^{\rm real} + \delta_3 & = & \frac{4g^2N_C}{(2\pi)^{D-1}}     
\left[\:\int \: \frac{d^{D-2}k}{\mbox{\boldmath $k$}^2} \:\int_{\mbox{\boldmath    
$k$}^2/s_{\Lambda}}^{\beta_0} \frac{d\beta}{2\beta}\right]_{\rm singular}     
\nonumber \\   
& & \nonumber \\     
& = & 4\bar g ^2\left(\frac{1}{\epsilon^2} +\frac{1}{\epsilon}     
\ln(s_{\Lambda}\beta_0)\right).     
\label{eq:a86}      
\eea   
We see that the sum of (\ref{eq:a84})-(\ref{eq:a86}) is zero, which means that we    
have verified the infrared safety of impact factors in general. \\         
     
\noindent {\large {\bf 7.~~Discussion}}     
     
In this paper we have investigated, to the next-to-leading order      
accuracy, the infrared properties of impact factors appearing in the      
BFKL description of small $x$ processes. We have shown, by explicit calculation, the    
cancellation of the infrared singularities in the impact factor of the virtual photon.      
The virtual photon impact factor was considered because of the importance of    
corrections to LLA for      
diffractive $q\bar q$ electroproduction and because of the possibility of      
calculating this impact factor in perturbation theory for large enough      
photon virtuality. We also presented a general proof of the      
infrared safety of the impact factors describing the transitions between      
colourless particles.      
       
We have considered impact factors in the general case of non-forward      
scattering and have used the definition of  impact factors      
given in \cite{NFS,FS2}. These impact factors are  defined      
for the case when the energy scale $s_0$ (in the Mellin      
transform factor $(s/s_0)^{\omega}$ in (\ref{eq:a2})) is a constant independent of the      
integration variables (the Reggeon \lq\lq masses"). One may prefer to use a scale    
which is dependent on these variables.   (For instance in the particular case of forward      
scattering, i.e.~for calculation of cross sections, the scale $\sqrt{\mbox{\boldmath    
$q$}_1^2\mbox{\boldmath $q$}_2^2}$ was used in \cite{NLA}).      
It was shown in \cite{A9} that to the NLA accuracy one can indeed change the scale    
$s_0$ in (\ref{eq:a2}) for any factorizable scale $f_1f_2$, with $f_i$ depending on    
$\mbox{\boldmath $q$}_i$ (and on $\mbox{\boldmath $q$}$), without changing the    
Green function, provided that the impact factors are also changed according to      
\bea   
\Phi _{A^{\prime }A}(\mbox{\boldmath $q$}_R,\mbox{\boldmath $q$};s_0)     
& \rightarrow & \Phi _{A^{\prime }A}(\mbox{\boldmath $q$}_R,\mbox{\boldmath     
$q$};s_0)  \nonumber \\   
& & \nonumber \\   
& & + \;\frac{1}{2} \:\int \:\frac{d^{D-2}q_{R^{\prime }}}{\mbox{\boldmath      
$q$}_{R^{\prime }}^2\mbox{\boldmath $q$}_{R^{\prime }}^{\prime 2}}\:\Phi     
_{A^{\prime }A}^{(0)}(\mbox{\boldmath $q$}_{R^{\prime }}, \mbox{\boldmath      
$q$})\:K^{(0)}(\mbox{\boldmath $q$}_{R^{\prime }}, \mbox{\boldmath     
$q$}_R;\mbox{\boldmath $q$}) \ln \left( \frac{f_R}{s_0}\right),      
\label{eq:a87}     
\eea   
where $K^{(0)}$ is the total non-forward LLA kernel,      
\begin{equation}     
K^{(0)}(\mbox{\boldmath $q$}_{R^{\prime }},      
\mbox{\boldmath      
$q$}_R;\mbox{\boldmath $q$})\:=\:     
 K_r^{(0)}(\mbox{\boldmath $q$}_{R^{\prime }},     
\mbox{\boldmath      
$q$}_R;\mbox{\boldmath $q$})\:       
+\;\mbox{\boldmath $q$}_{R^{\prime }}^{\prime 2}\mbox{\boldmath      
$q$}_{R^{\prime }}^2\:\delta (\mbox{\boldmath $q$}_R-\mbox{\boldmath      
$q$}_{R^{\prime }})\: \left( \omega (t_R)\:     
+\:\omega (t_R^{\prime }) \right).   
\label{eq:a88}     
\end{equation}      
$K_r^{(0)}$ is the part of the LLA kernel related to real      
gluon production, (\ref{eq:a20}), and $\omega (t)$ is the gluon      
Regge trajectory. The properties of      
the impact factors (\ref{eq:a17a}) and the infrared safety      
of the total kernel (due to cancellation of the infrared      
singularities connected with  $ K_r^{(0)}$ and with the      
gluon trajectory) guarantees that the transformation       
(\ref{eq:a87}) does not change the infrared properties of      
the impact factors.      
     
We emphasize that the above definitions differ from those introduced,       
for the case of the forward scattering, in \cite{C} and used in       
\cite{CC} and in \cite{CCC},  where      
impact factors of partons (quarks and gluons) were calculated for a colour singlet state    
in the $t$ channel. The definition advocated in these papers is based on \lq\lq the  
requirement that the subtracted leading term should satisfy at finite energies      
for the impact factors to be well defined (e.g., without spurious      
infrared divergencies)" \cite{CCC}. In order to fulfil this requirement   
additional operator factors $H_L, H_R$ were introduced \cite{CC},    
\cite{CCC} in the Green function $G_{\omega}$.  In our opinion, the only   
physical requirement is the  absence of infrared singularities in the   
impact factors of colourless objects, and we have explicitly demonstrated here   
that our definition of the impact factors fulfills this important requirement. \\   
   
\noindent {\large \bf Acknowledgements}   
   
We thank the Royal Society and INTAS (95-311) for financial support.

\newpage   
\noindent {\large \bf Figure Captions}   
\begin{itemize}   
\item[Fig.~1] Schematic representation of the process $A + B \rightarrow A^\prime +    
B^\prime$ displaying the structure of (\ref{eq:a2}).  The zig-zag lines represent    
Reggeized gluon exchange.   
   
\item[Fig.~2] The intermediate process $A + B \rightarrow a + b$ mediated by    
Reggeized gluon exchange.   
   
\item[Fig.~3] The lowest-order Feynman diagrams for the process    
$\gamma^* q \rightarrow (q\bar{q})q$, showing the particle four momenta.   
   
\item[Fig.~4] Schematic representation of the vertex $\Gamma_{q\bar{q},    
\gamma^*}^{(0) i}$, showing, at lowest order, the transition $\gamma^*R \rightarrow    
q\bar{q}$.  
  
\item[Fig.~5] The two-gluon exchange Feynman diagrams mediating the process    
$\gamma^* q \rightarrow (q\bar{q}) q$.   
   
\item[Fig.~6] Feynman diagrams mediating $\gamma^* q \rightarrow (q\bar{q})q$    
which must be considered with those of Fig.~5.   
   
\item[Fig.~7] Diagrams containing the gluon-gluon-Reggeon vertex    
which contain the corrections to the $\Gamma_{q\bar{q},    
\gamma^*}$ coming from Figs.~5 and 6.  
   
\item[Fig.~8] The gluon-gluon-Reggeon vertex of (\ref{eq:a35}).   
   
\item[Fig.~9] Loop diagrams within the produced $q\bar{q}$ pair.  The quark    
(antiquark) self-energy diagrams are not shown.   
   
\item[Fig.~10] The real gluon emission diagrams from the $\Gamma_{q\bar{q},    
\gamma^*}$ vertex.   
\end{itemize}       
     
\newpage
\begin{center}
\vspace*{-17cm}
\epsfig{figure=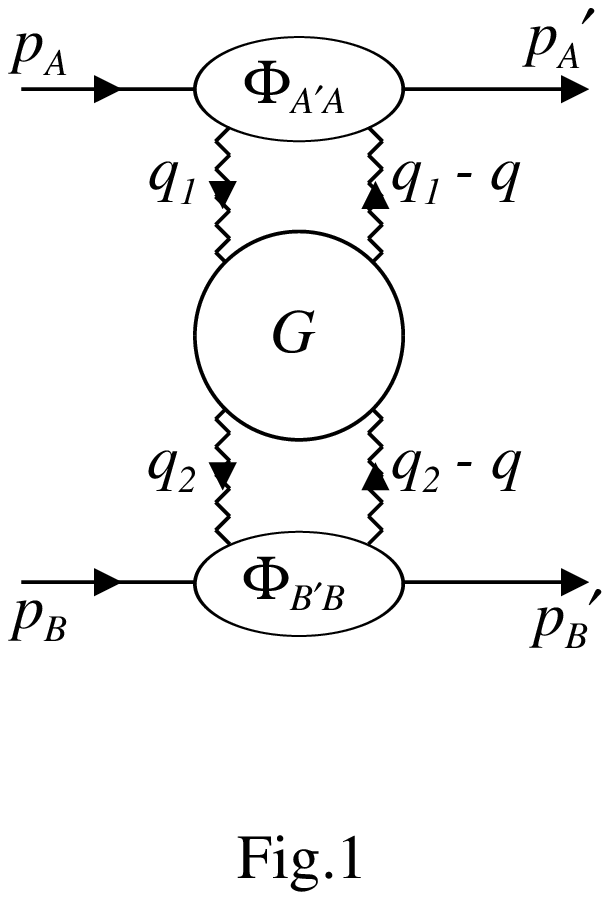,width=8cm}\\
\vspace{3cm}
\epsfig{figure=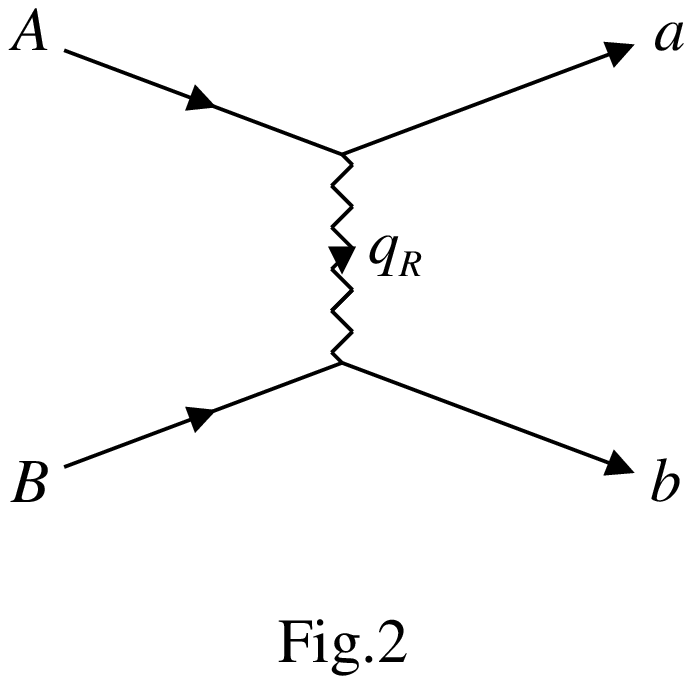,width=7cm}
\newpage
\epsfig{figure=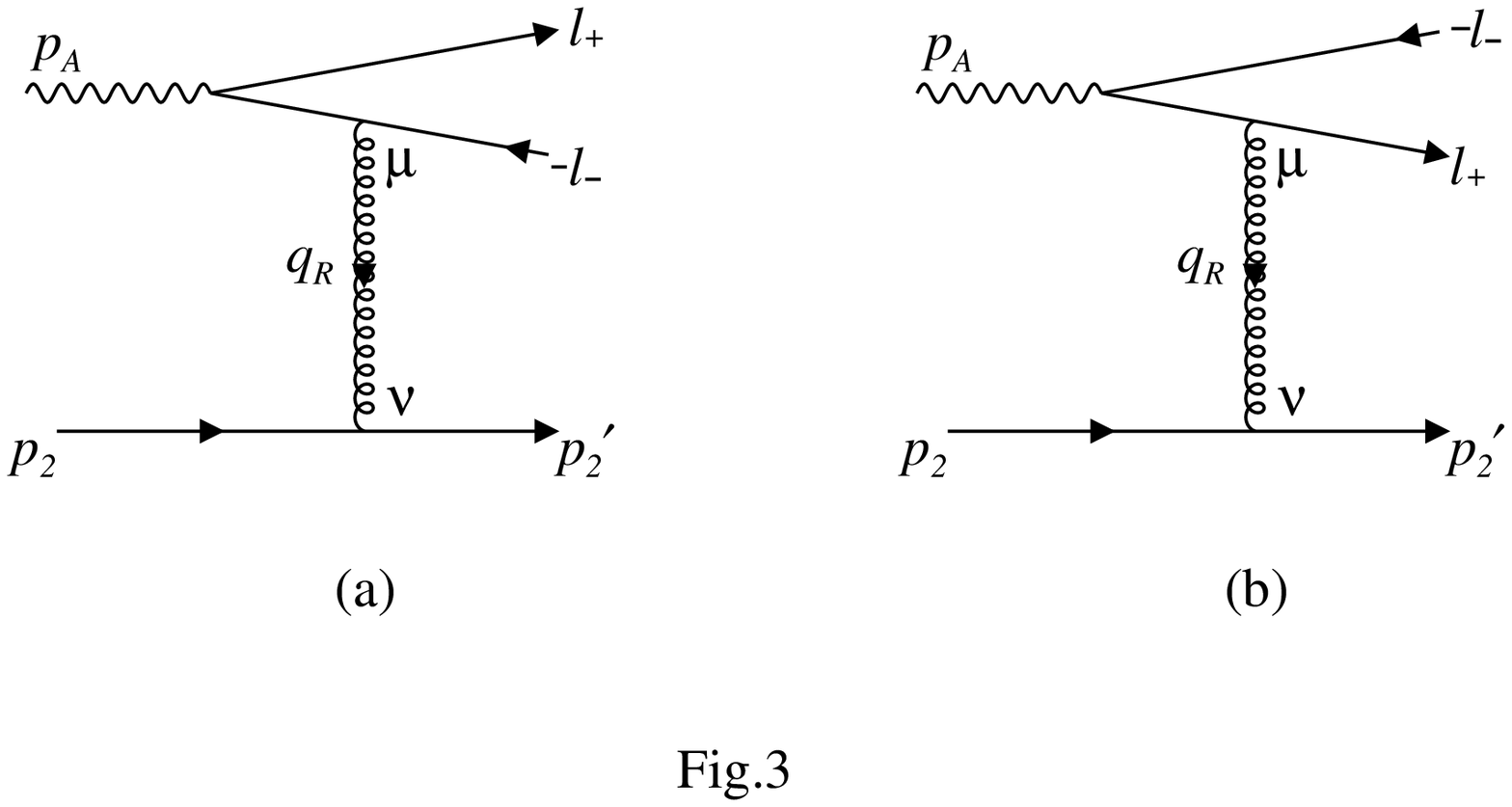,width=16cm}\\
\vspace*{3cm}
\epsfig{figure=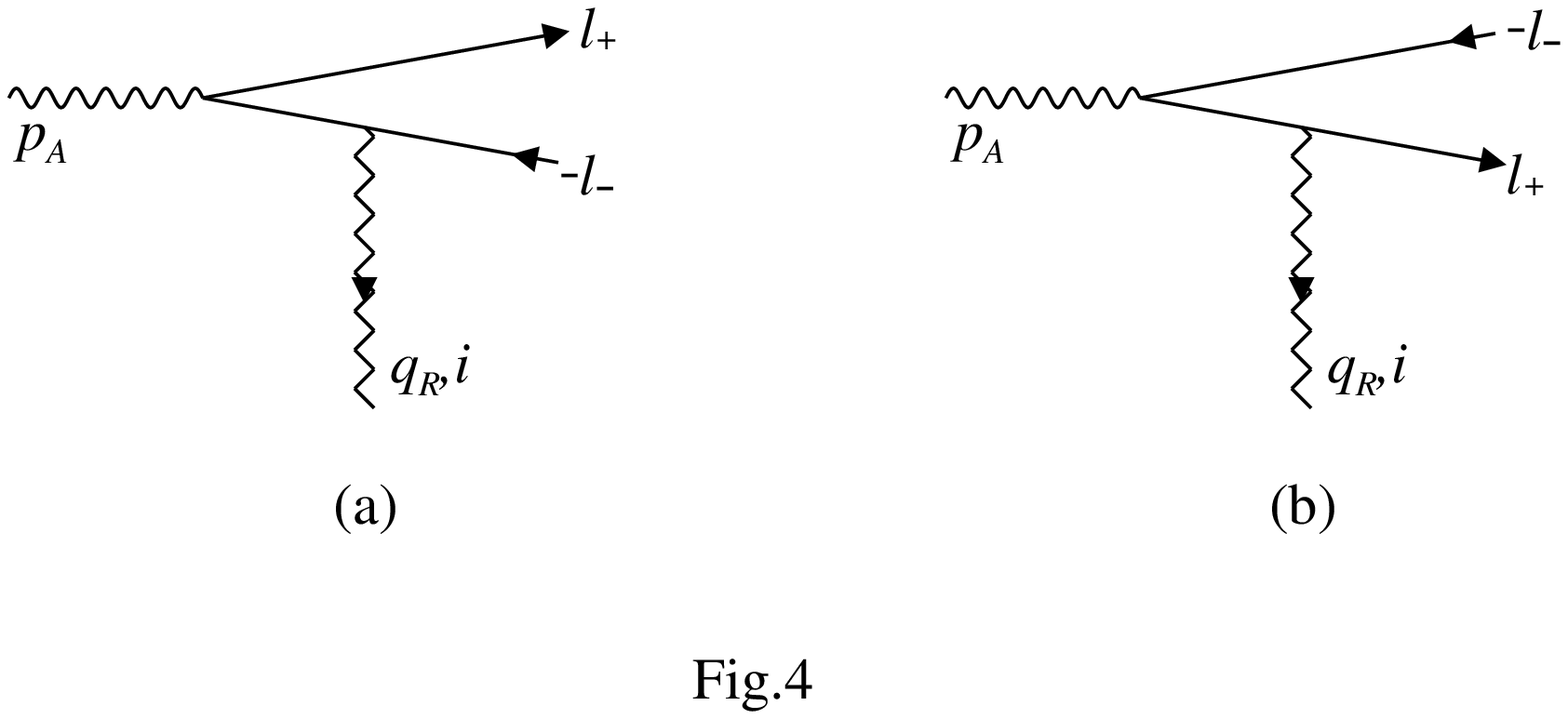,width=16cm}
\newpage
\epsfig{figure=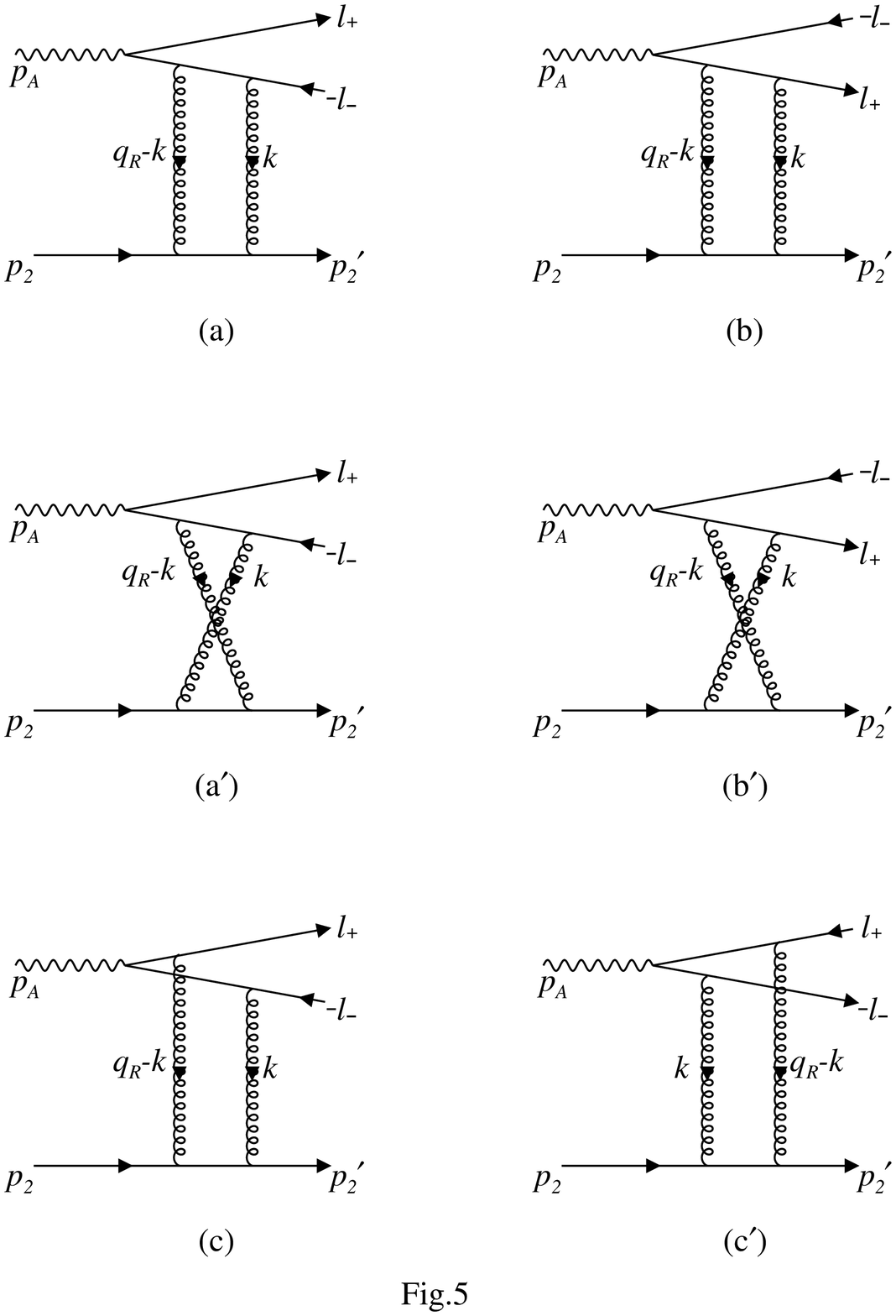,width=16cm}
\newpage
\epsfig{figure=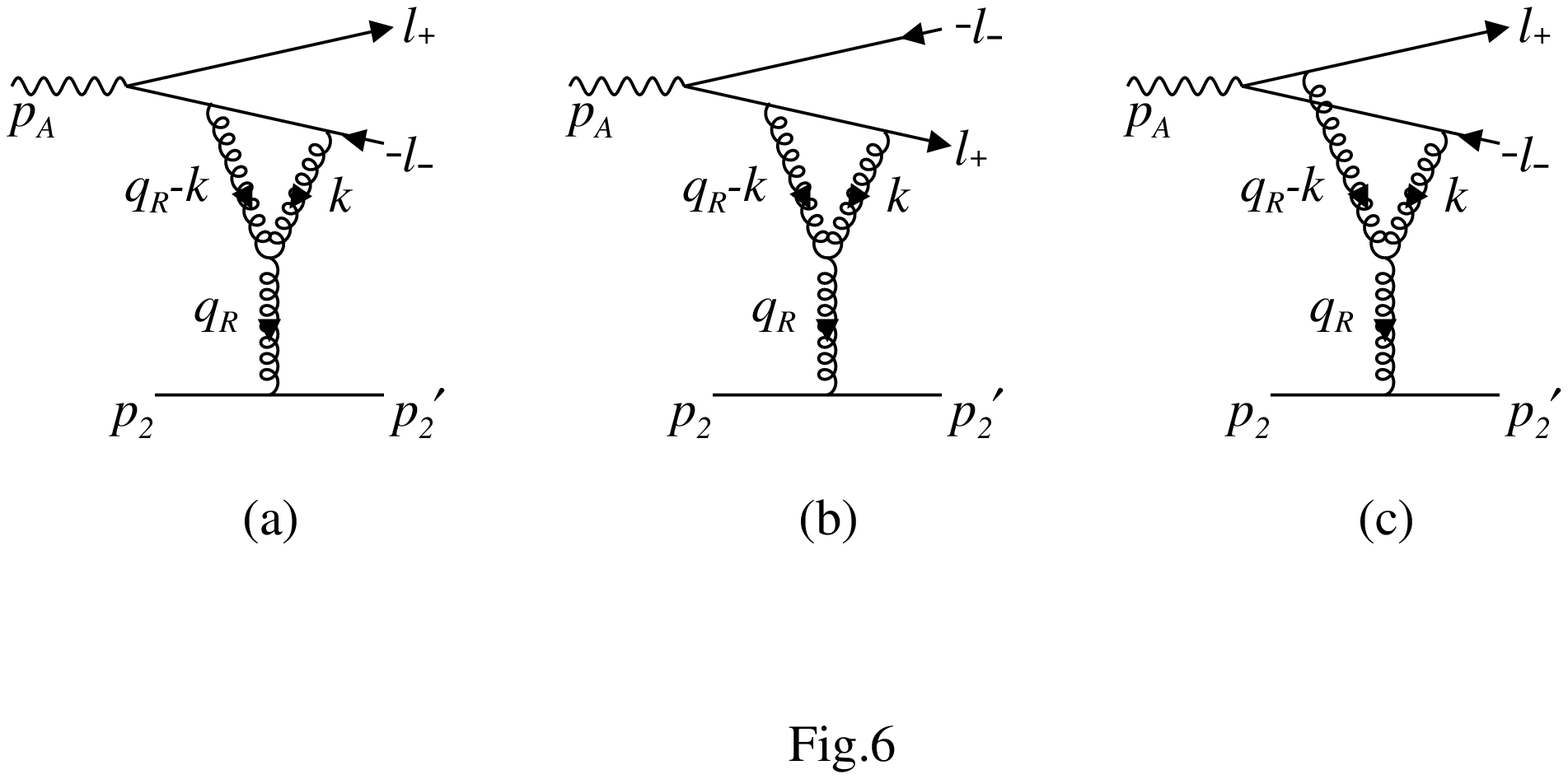,width=16cm}\\
\vspace*{3cm}
\epsfig{figure=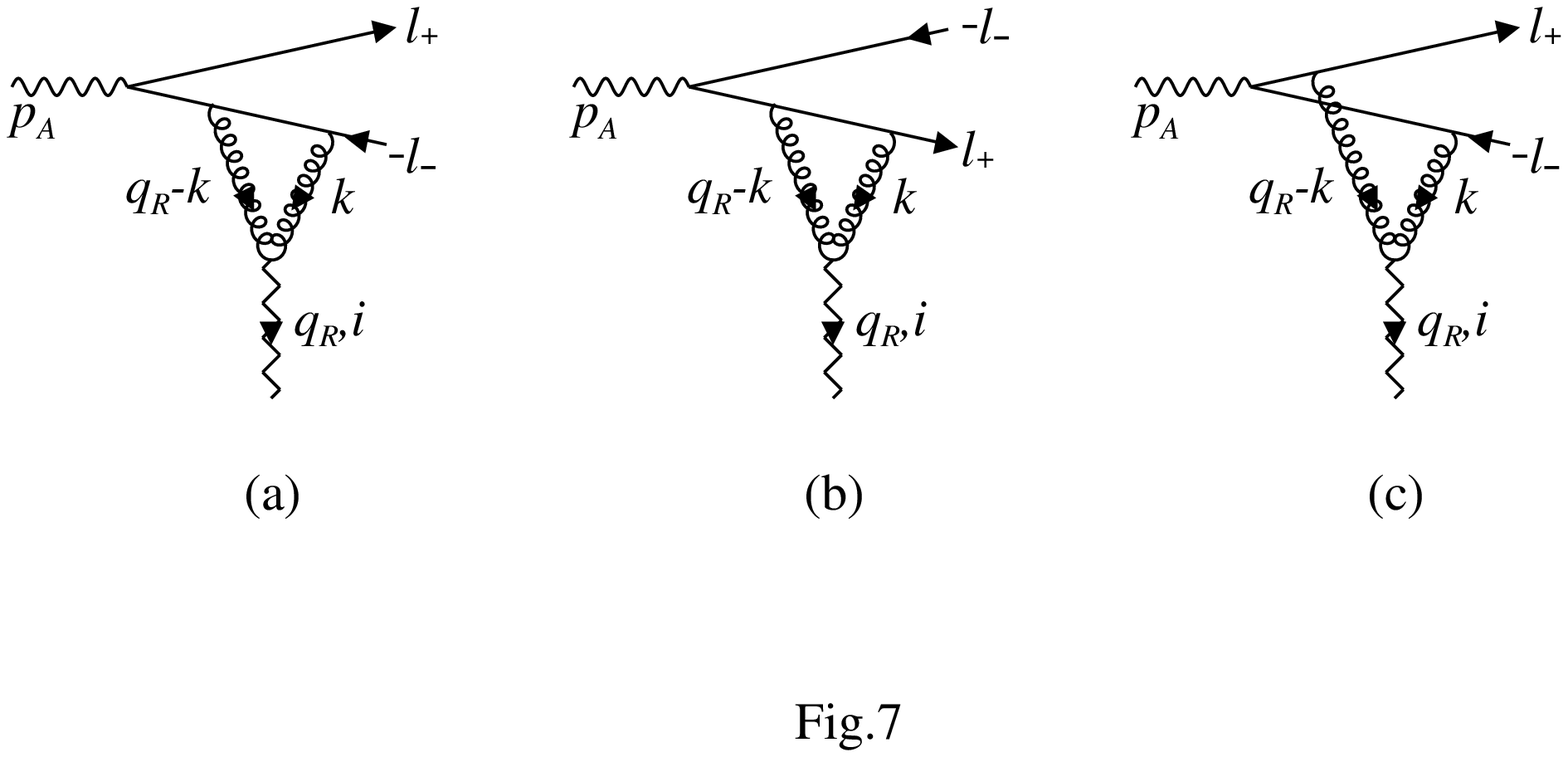,width=16cm}
\newpage
\epsfig{figure=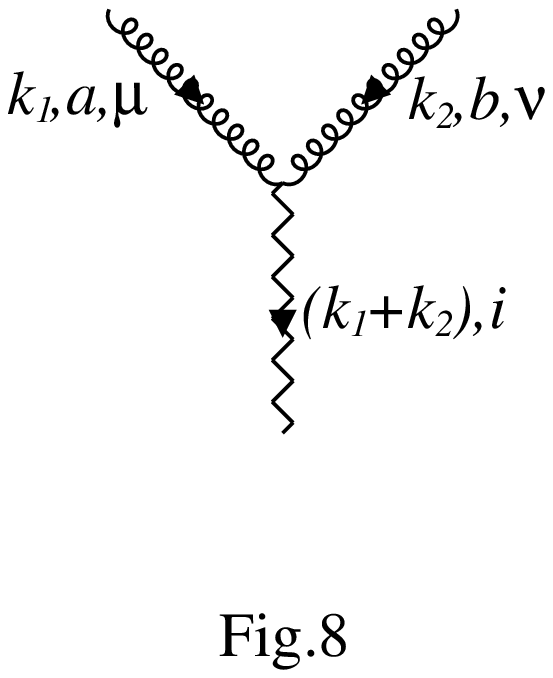,width=5cm}\\
\vspace*{3cm}
\epsfig{figure=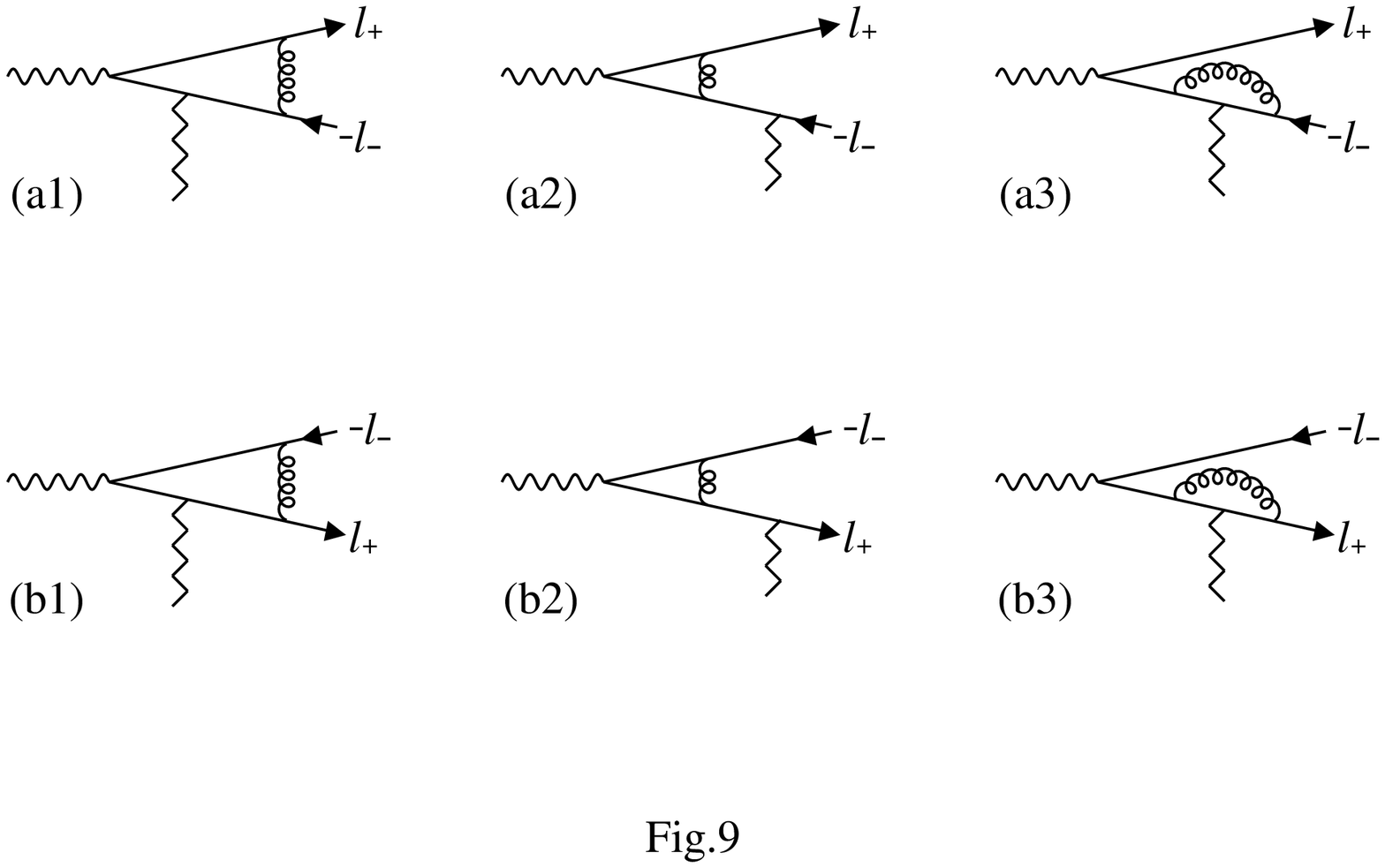,width=16cm}
\newpage
\epsfig{figure=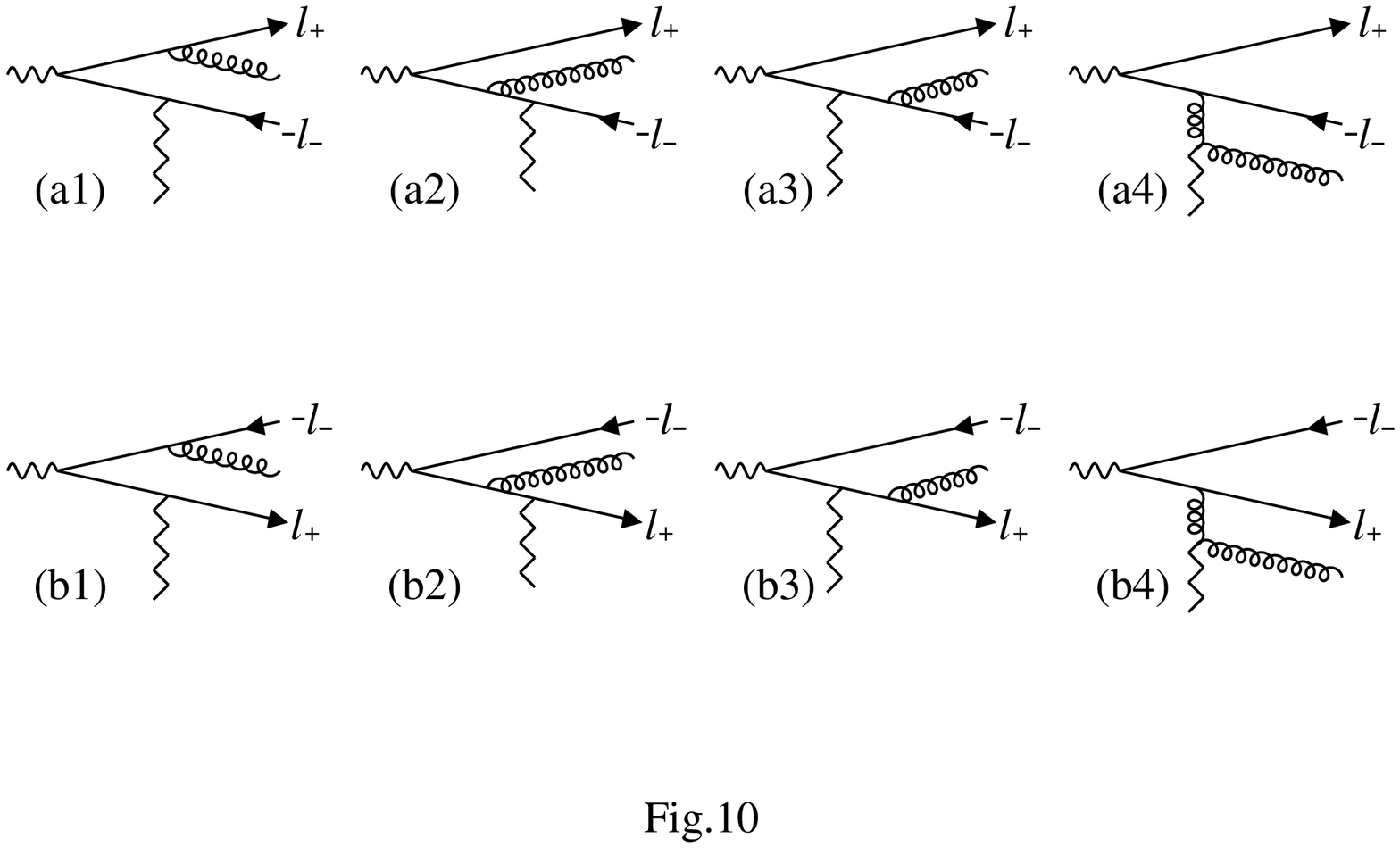,width=16cm}
\end{center}     
\end{document}